# Flexible Wideband Filtering Monopole Antenna With Stable High Omnidirectionality

Runkai Song, *Graduate Student Member, IEEE*, Fan Qin, *Member, IEEE*, Chao Gu, *Member, IEEE*, Jiangzhou Wang, *Fellow, IEEE*, Wenchi Cheng, *Senior Member, IEEE*, and Steven Gao, *Fellow, IEEE*

***Abstract*——This article presents a wideband flexible filtering monopole antenna with symmetric structure for stable high omnidirectionality. It is based on a monopole antenna, which is printed on a single-layer flexible substrate. Two folded parasitic strips with different length are devised on both sides of the driven monopole, giving filtering responses in the higher and lower band without filtering circuits. Since the asymmetric filtering structure adversely affects in-band omnidirectionality, this baseline design is extended with symmetric filtering structure to improve omnidirectionality and bandwidth. In the proposed design, a pair of parasitic strips are devised on the both sides of monopole antenna symmetrically, achieving a radiation null in the higher band. Then, by loading a pair of folded parasitic strips on the both sides of feed line with slotted metal ground, a radiation null is realized in the lower band. Besides, the driven monopole is slotted symmetrically for wideband operation. By adopting a fully symmetric filtering structure, the proposed design effectively suppresses the impact of the parasitic elements on the in-band omnidirectional radiation pattern, thereby achieving high omnidirectionality. Furthermore, the proposed antenna exhibits stable performance under different bending radii. To verify our design concept, an antenna prototype is fabricated. Both the flat and bent antennas are measured. The results show that the proposed antenna has a -10 dB impedance bandwidth of 45.6%, an in-band gain about 2 dBi, and an out-of-band radiation suppression more than 11 dB. The measured omnidirectionality has variations less than 0.8 dB without bending and 1 dB with a bending radius of 30 mm. This design offers several advantages including stable high omnidirectionality across a wide bandwidth, flexible conformal capability, and filtering property.**

## I. Introduction

In the modern wireless systems, omnidirectional antennas serve as a fundamental component, providing large-signal coverage with uniform radiation patterns. Their application scenarios span key areas [1]-[4], including Internet of Things (IoT) deployments, integrated sensing and communication (ISAC), in-door base stations, and wireless local area network (WLAN). In these systems, omnidirectional antennas are expected to exhibit high omnidirectionality to provide consistent signal coverage for all users. Various techniques have been mentioned for omnidirectional radiation patterns, such as dipole antennas [5]-[7], monopole antennas [8]-[11], slot antennas [12]-[14], and dielectric resonator antennas [15]-[16]. With the properties including compact size, small cross-polarization, broad impedance band, and easy fabrication, monopole antennas are extensively used to achieve omnidirectional radiation. Moreover, integrating filter and antenna as two vital elements in the radio frequency front into antenna design, research on filtering antennas has received growing attention due to their advantages of low insertion loss and small size [17]-[21]. Some omnidirectional filtering antenna design has been proposed [22]-[25]. In [22], an inductive window bandpass filter and a planar coaxial collinear radiation element are connected for linearly polarized omnidirectional filtering antenna design. A compact tunable band-notched filtering antenna reported in [23] is achieved by incorporating an ultra-wideband filter on its feed line. The above designs introduce extra filtering networks to achieve omnidirectional filtering property, increasing the size and insertion loss of those filtering designs to some extent.

Recently, the omnidirectional filtering antenna structures without extra filtering circuits have been introduced [26]-[31]. In [26], two parasitic strips are attached on a folded dipole to introduce the radiation nulls for omnidirectional filtering antenna design. Short pins and metallic vias are employed to improve the frequency selectivity, as reported in [27] and [28]. Moreover, an omnidirectional filtering design introduced in [29] is achieved by connecting three loops to produce three tunable radiation nulls for frequency selectivity. In [30], nonradiative elements including a coupled U-shaped microstrip line and two I-shaped slots are added to the feeding network for filtering responses. The filtering antenna design reported in [31] achieves stable omnidirectional radiation patterns by introducing a symmetrical folded patch architecture. These antennas have good filtering capability with integrated filtering structures, achieving more compact size, higher integration and lower insertion loss.

Besides, the combination of thin profile, compact size, conformal capability, and flexibility has improved the development of flexible antenna designs [32]-[35]. Most recently, integrating flexibility and filtering capability, some flexible filtering antennas have been reported [36]-[37]. The

Runkai Song, Fan Qin, and Wenchi Cheng are with the School of Telecommunications Engineering, Xidian University, Xi’an 710071, China (e-mail: srk@stu.xidian.edu.cn; fqin@xidian.edu.cn; wccheng@xidian.edu.cn).

Chao Gu is with ECIT Institute, Queen’s University Belfast, BT39DTBelfast, U.K. (e-mail: chao.gu@qub.ac.uk).

Jiangzhou Wang is with the School of Engineering and Digital Arts, University of Kent, Canterbury CT2 7NT, U.K (e-mail: j.z.wang@kent.ac.uk).

Steven Gao is with the Department of Electronic Engineering, Chinese University of Hong Kong, Hong Kong (e-mail: scgao@ee.cuhk.edu.hk).

dual-band flexible filtering endfire antenna reported in [36] achieves integrated filtering design with an ultra-low-profile flexible structure. In [37], a flexible wearable filtering antenna with stable performance after bending is realized by loading a pair of inverted F-shaped slots on the radiation patch to suppress non-radiated lateral current components. These flexible filtering antennas achieve stable and effective out-of-band radiation suppression after bending. However, the applications of integrated filtering structure, such as parasitic strips, short pins and slots, will affect the in-band omnidirectionality to some extent. Most existing studies only discuss the out-of-band suppression of integrated structures, neglecting their influence on in-band omnidirectional radiation. Therefore, it is necessary to discuss how to mitigate the impact of parasitic elements for high omnidirectionality. Besides, the existing flexible filtering antennas have large size, relatively narrow bandwidth, and restricted working bending radius, which can be further improved for better performance.

In this article, a wideband flexible omnidirectional filtering antenna with high omnidirectionality is proposed. The design is based on a monopole antenna, which is printed on a single-layer flexible substrate. A pair of parasitic strips are devised on both sides of the driven monopole symmetrically, introducing a radiation null in the upper band. Then, by adding a pair of folded parasitic strips around the feed line with slotted metal ground, a radiation null can be achieved in the lower band. Besides, symmetric slots are incorporated into the driven monopole for wide operating band. Due to the implementation of a perfectly symmetric filtering antenna structure, the proposed antenna exhibits high omnidirectionality in the wide operating band. Besides, the antenna can maintain stable bandwidth, high omnidirectionality, and effective frequency selectivity under different bending radii. To verify this design, the proposed antenna is designed and fabricated. The working mechanism of the proposed antenna is analyzed in detail. The design guidelines and measured results are also presented.

This paper presents three main innovations. Firstly, a novel fully symmetric omnidirectional filtering antenna structure based on printed flexible monopole is proposed. By effectively suppressing the interference of parasitic elements on in-band omnidirectional radiation, the proposed antenna achieves a high omnidirectionality compared with the existing filtering design. Besides, the bandwidth of the proposed antenna is further improved compared to previously published wideband omnidirectional filtering antennas. Moreover, the proposed design is the first to realize a flexible filtering antenna with stable omnidirectional radiation. It has compact size, broader bandwidth, and greater working bending curvature compared with the existing flexible filtering design.

## II. Antenna Design And Analysis

### A. *Baseline filtering structure based on printed monopole*

The typical electric fields of a quarter-wave monopole can be calculated. The electric fields of the monopole are stronger near the area of two ends of the monopole radiator. The electric coupling will be excited when a parasitic strip is placed near the two end parts of the monopole. As discussed in [26], the cross-coupling theory points out that the electric coupling can introduce transmission zero. Therefore, we speculate whether the electrical coupling properties of the printed monopole can be utilized to design an omnidirectional filtering antenna structure. Parasitic strips whose resonant frequency is at the lower or higher band of the driven monopole can be introduced, exciting radiation nulls for filtering capability. To suppress out-of-band radiation effectively, two radiation nulls working at the lower band and higher band are required.

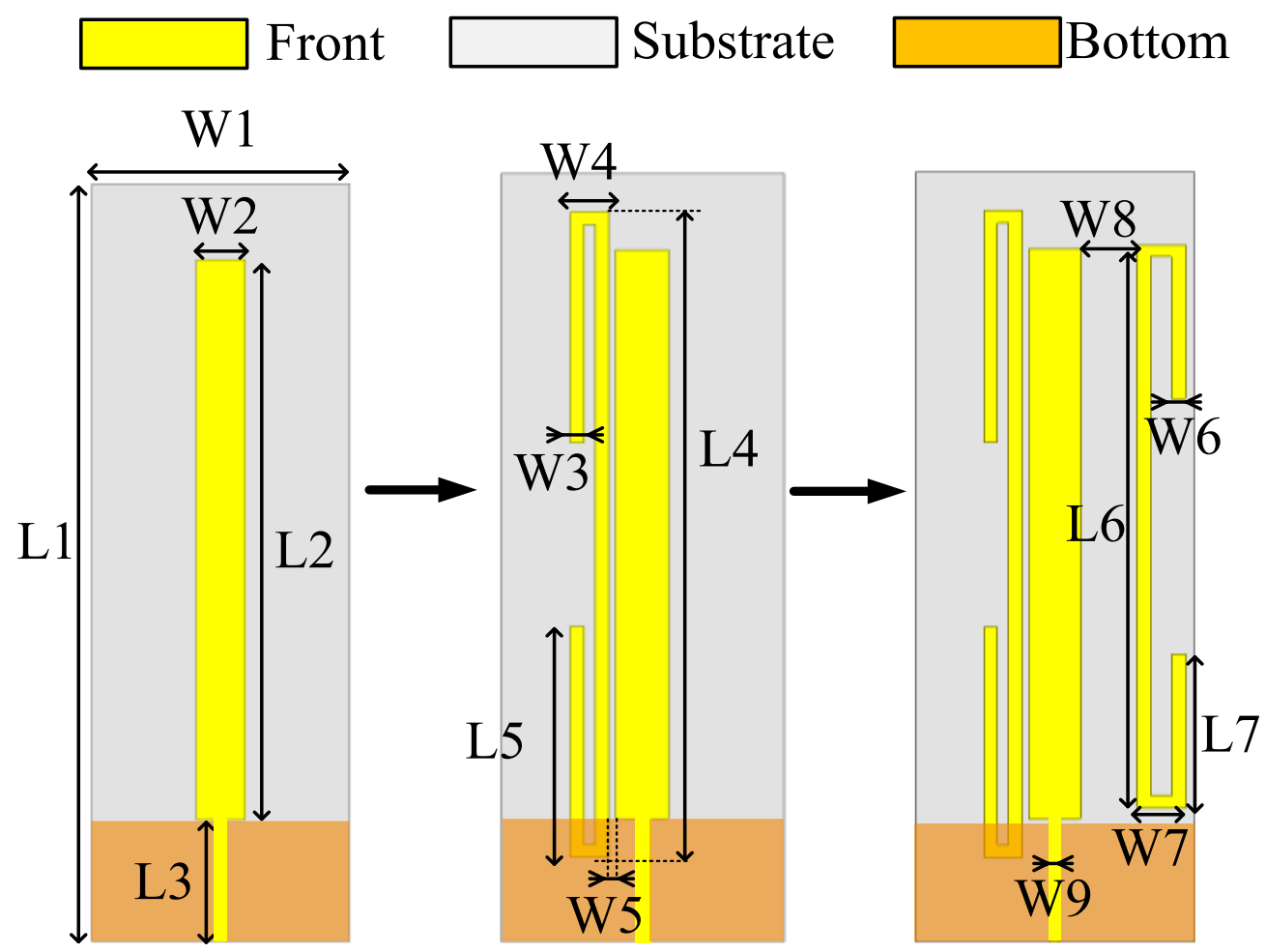

Fig. 1. Structure evolution with adding two asymmetric folded parasitic strips on a driven monopole for filtering capability. (L1=50 mm, L2=37 mm, L3=8 mm, L4=42 mm, L5=15 mm, L6=36.6 mm, L7=10 mm, W1=16 mm, W2=3 mm, W3=0.8 mm, W4=2.2 mm, W5=0.4 mm, W6=0.8 mm, W7=2.5 mm, W8=3.2 mm, W9=0.21 mm.)

To verify this method, two parasitic strips with different lengths are attached on a printed monopole as shown in Fig. 1. The filtering monopole is printed on a single-layer polyimide (PI) flexible substrate. Fig. 2 shows the simulated realized gain and S11 results for the driven monopole, monopole with a longer strip and baseline design. By adding a folded longer strip on the driven monopole, its operating bandwidth ranges from 2.1 GHz to 2.4 GHz as shown in the Fig. 2(a). Besides, a radiation null is observed in the lower band as illustrated in Fig. 2(b), achieving an out-of-band radiation suppression more than 20 dB. Moreover, by introducing a shorter folded parasitic strip on the other side of the driven monopole, the operating band of the baseline design is extended to 28% (from 2.1 GHz to 2.7GHz). A higher band radiation null is realized at 2.8 GHz with steep out-of-band gain reduction. After adding two parasitic strips, two radiation nulls can be observed at the lower and higher band, realizing effective out-of-band radiation suppression.

Following the filtering design of the baseline antenna, its in-band radiation characteristics deserves further discussion. The in-band radiation patterns of the baseline design and the driven monopole at 2.5 GHz are presented in Fig. 3. In the H-plane, the co-polarization fields of the driven monopole have variations less than 0.3 dB, while the omnidirectionality of the baseline design increases to 3.5 dB after adding two parasitic strips for filtering capability. The in-band omnidirectional radiation pattern is seriously interfered after adding asymmetric filtering strips, which leads to a significant deterioration in omnidirectionality. To further investigate how the parasitic strips affect in-band omnidirectionality, the in-band current distributions (at 2.5 GHz) for the driven monopole and the baseline design are shown in Fig. 4. Due to differences in the length, shape, and coupling distance of the parasitic strips, the path, intensity and direction of the in-band coupled currents on the strips are different. Therefore, the radiation generated by the coupled currents on the asymmetric parasitic strips is not omnidirectional in the far field, leading to distortion of the omnidirectionality. Besides, the current distributions on the driven monopole become unbalanced under the influence of coupling between the monopole and asymmetric strips on both sides, which results in a deterioration in its omnidirectionality. Therefore, how to further improve filtering structures to minimize their impact on the in-band omnidirectionality become a key challenge.

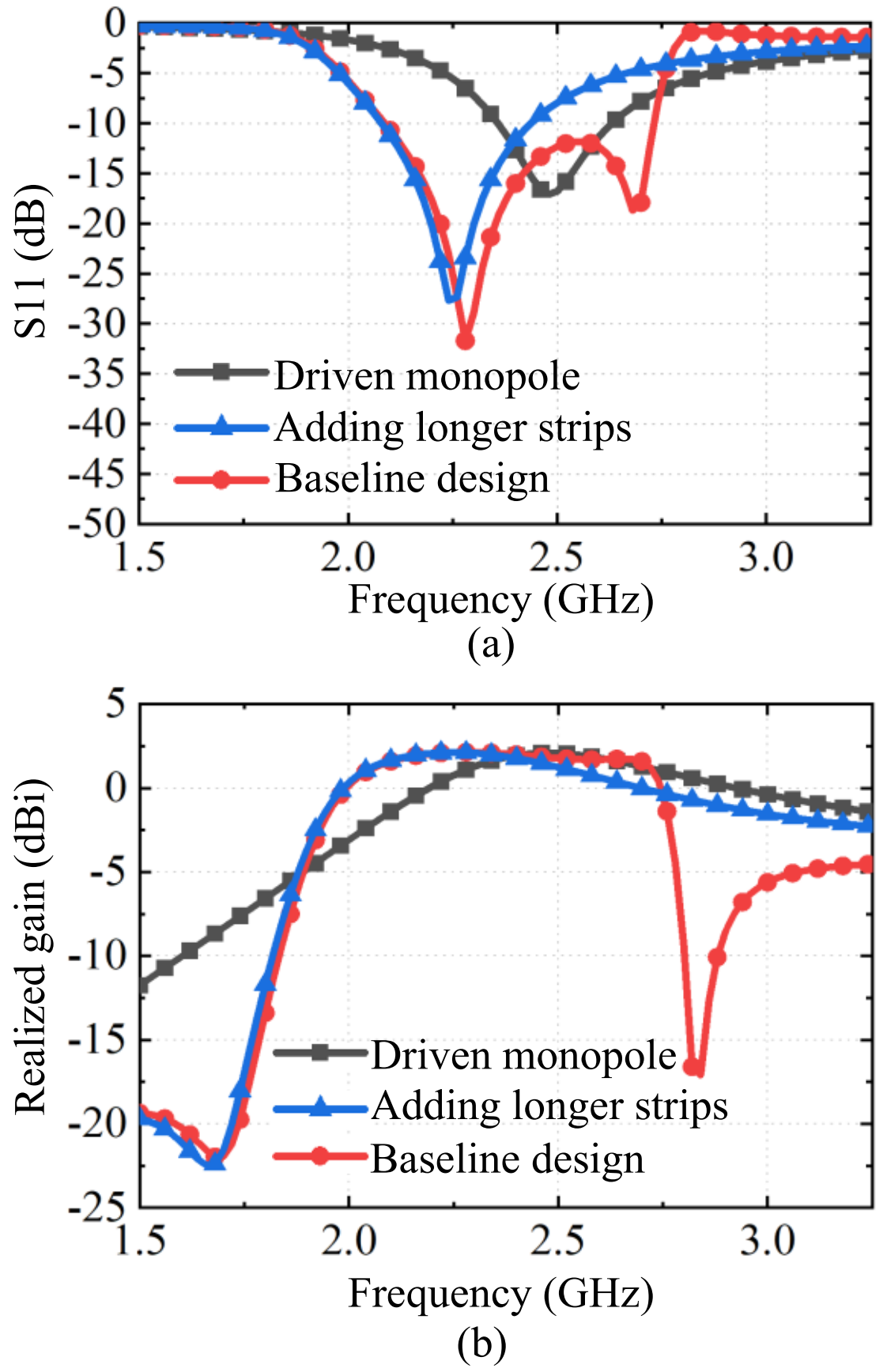

Fig. 2. Simulated S11 and realized gains results for the driven monopole, monopole with a longer strip and the baseline design.

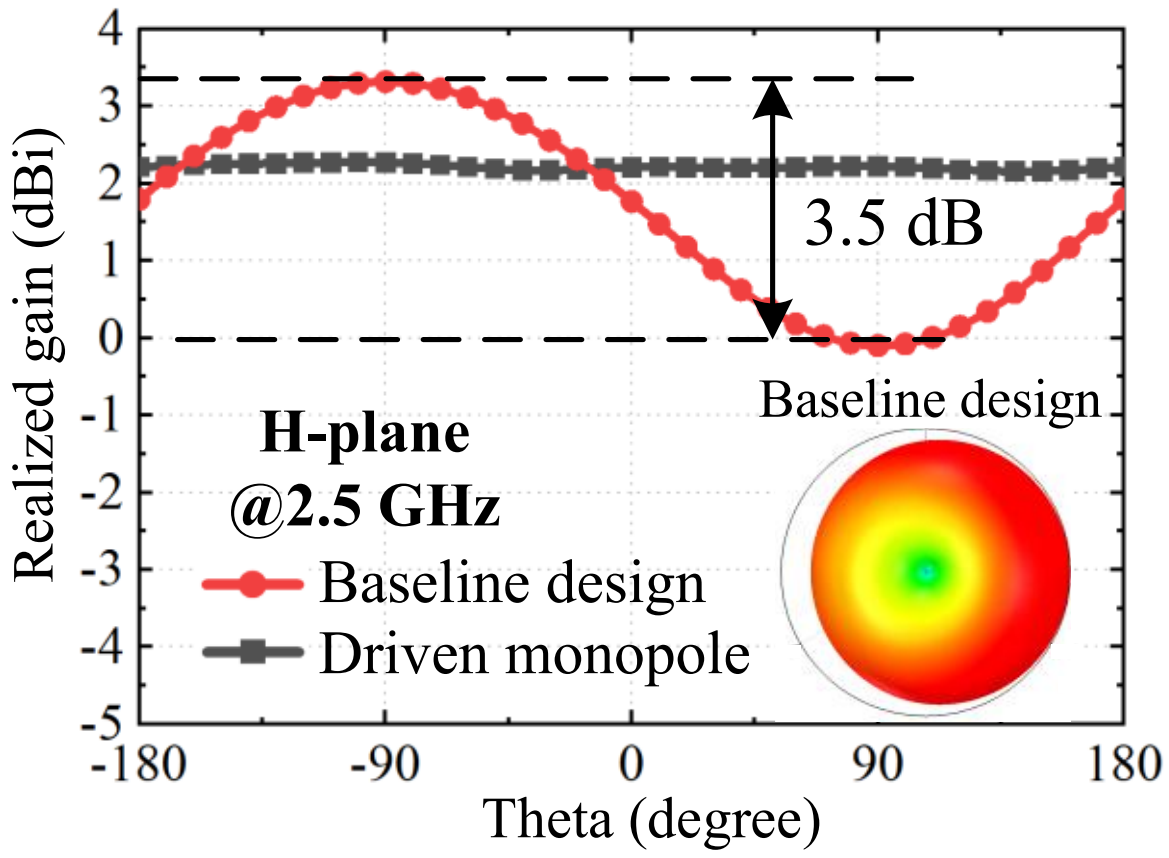

Fig. 3. In-band H-plane radiation patterns (at 2.5GHz) of the baseline design and the driven monopole.

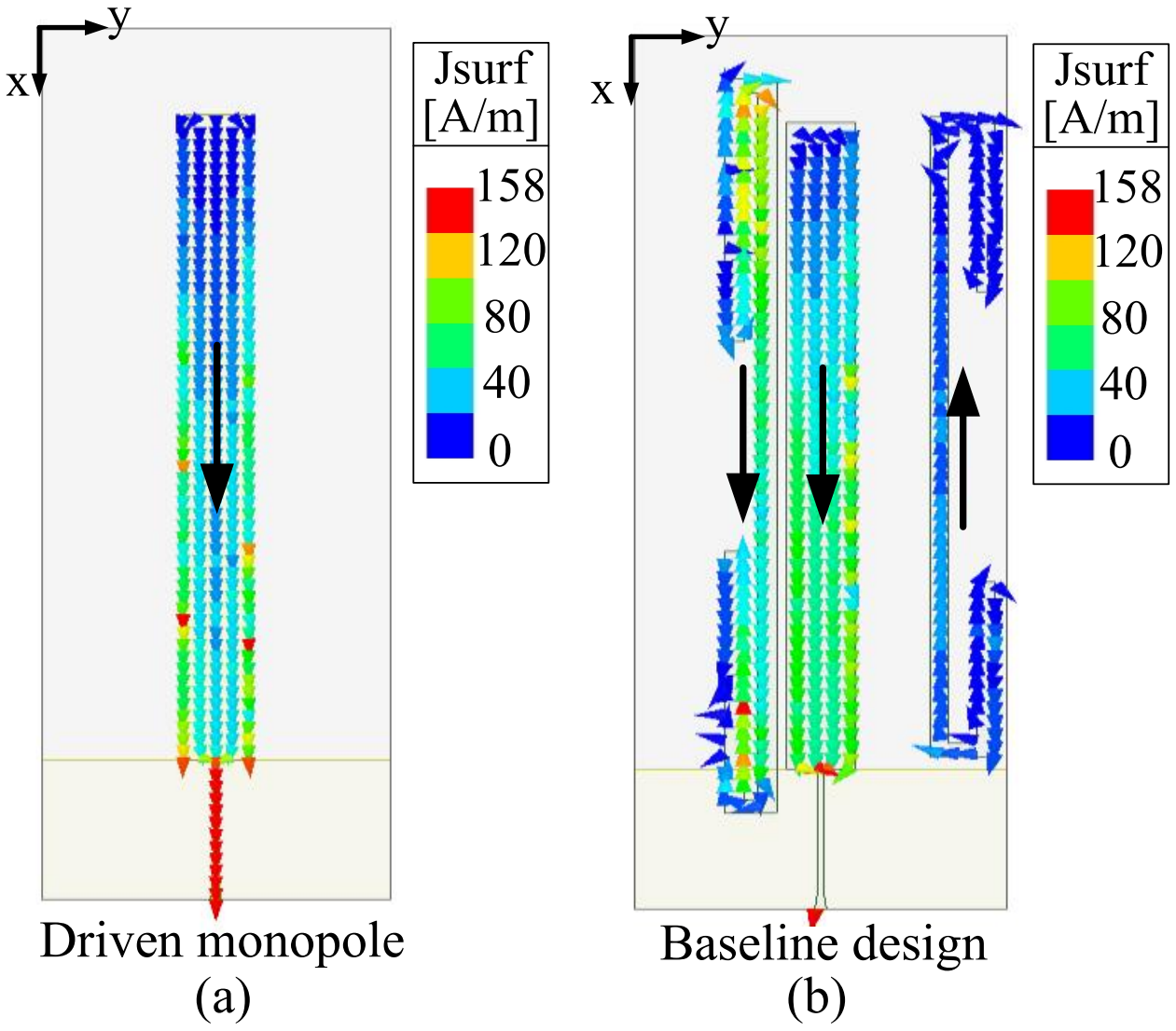

Fig. 4. Simulated in-band surface current distributions for the baseline design and the driven monopole at 2.5 GHz.

### *B. Improved Symmetric Filtering Antenna Structure for High Omnidirectionality and Wideband*

Owing to the introduction of parasitic strips with different lengths and coupling distances, the in-band omnidirectionality is critical interfered. To effectively mitigate the impact of filtering parasitic structures on radiation pattern, a symmetrical omnidirectional filtering antenna structure is proposed.

The symmetric filtering structure involution of the proposed antenna is shown in Fig. 5. Starting from the baseline antenna, Ant.1 replaces the asymmetrical folded parasitic strips on both sides of the monopole with fully symmetric parasitic strips of equal length. Fig. 6 presents the S11 and realized gain results for Ant. 1. As shown in the Fig. 6(a), Ant. 1 achieves a substantial bandwidth enhancement over the baseline design. Moreover, it can be observed from Fig. 6(b) that a radiation null is created in the higher band after adding symmetric linear parasitic strips. However, compared with the asymmetric baseline design, symmetric parasitic strips of equal length can introduce only one radiation null. Therefore, after achieving a higher band null, a pair of folded parasitic strips are introduced at both sides of the feed line with slotted metal ground from Ant. 1 to Ant. 2 shown in the Fig. 5. The S11 and realized gain results from Ant. 1 to Ant. 2 are introduced in Fig. 7. As can be seen from Fig. 7(b), Ant. 2 exhibits a new lower band radiation null, achieving out-of-band radiation suppression capability with two radiation nulls. However, as illustrated in Fig. 7(a), the S11 parameter of the Ant. 2 deteriorates from below -10 dB to approximately -5 dB from 2.5 GHz to 3.75 GHz. Consequently, the in-band realized gain of Ant. 2 exhibits a corresponding reduction.

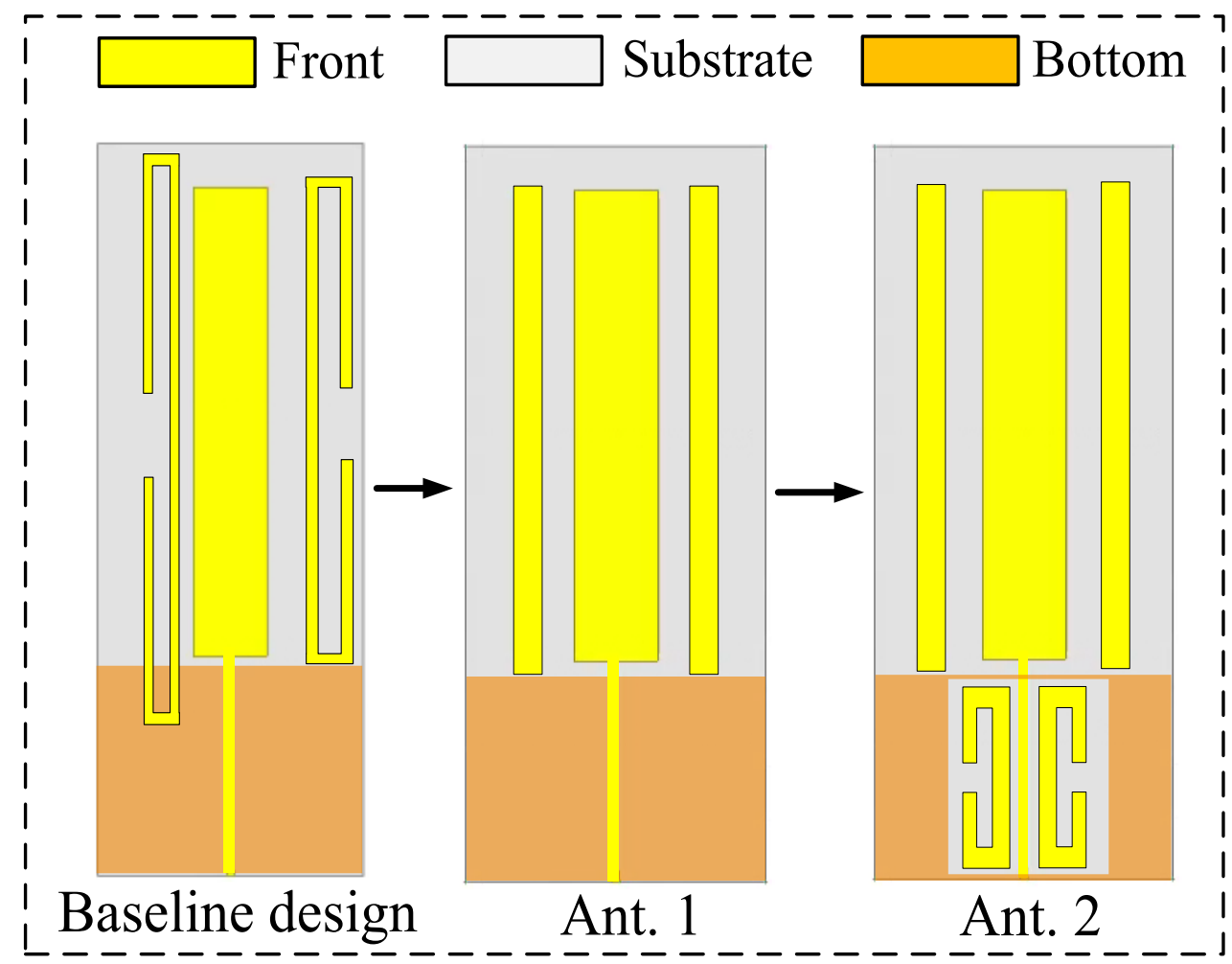

Fig. 5. Design procedure of the proposed symmetric filtering structures.

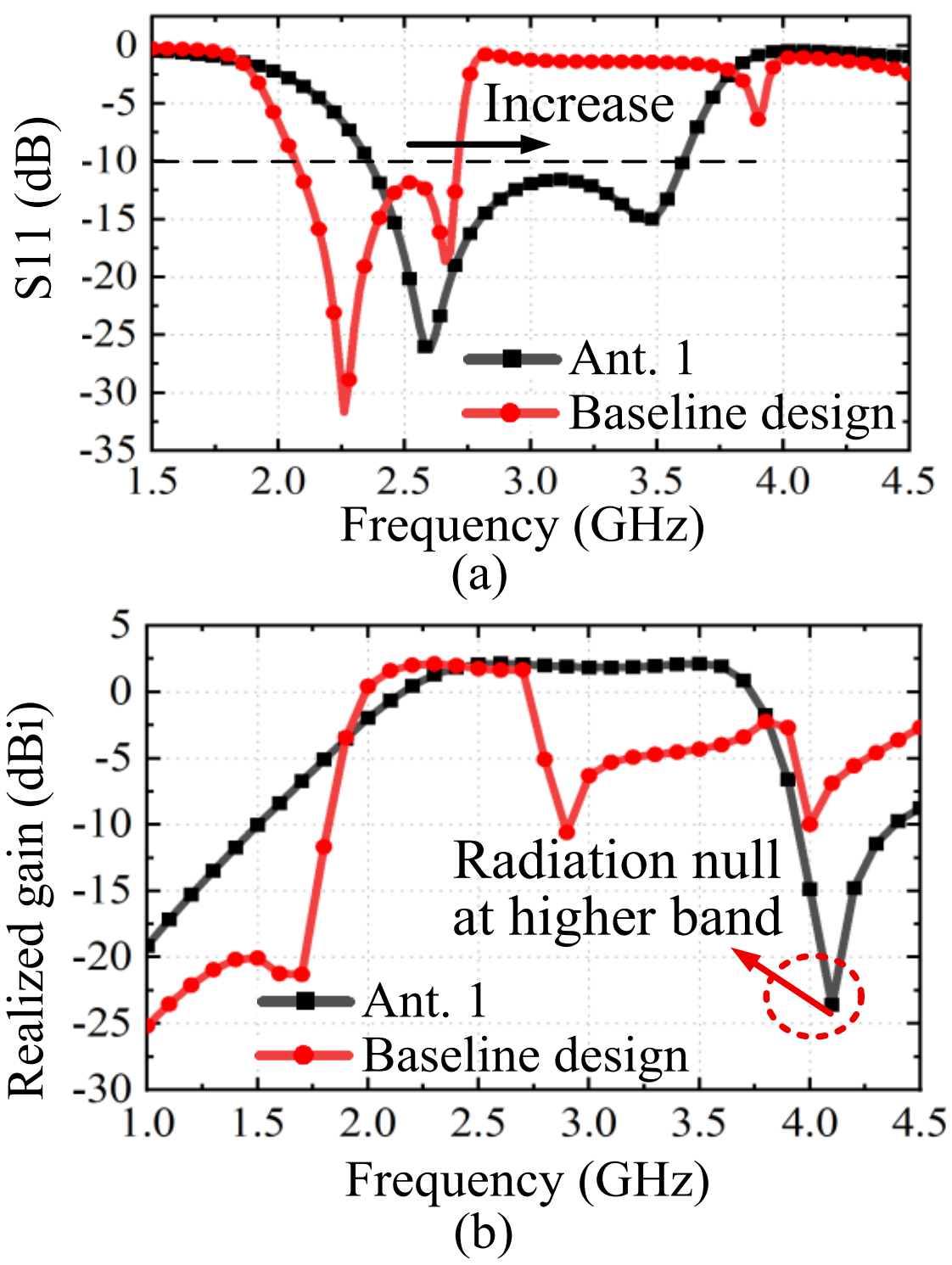

Fig. 6. S11 and realized gain results from the baseline design to Ant. 1.

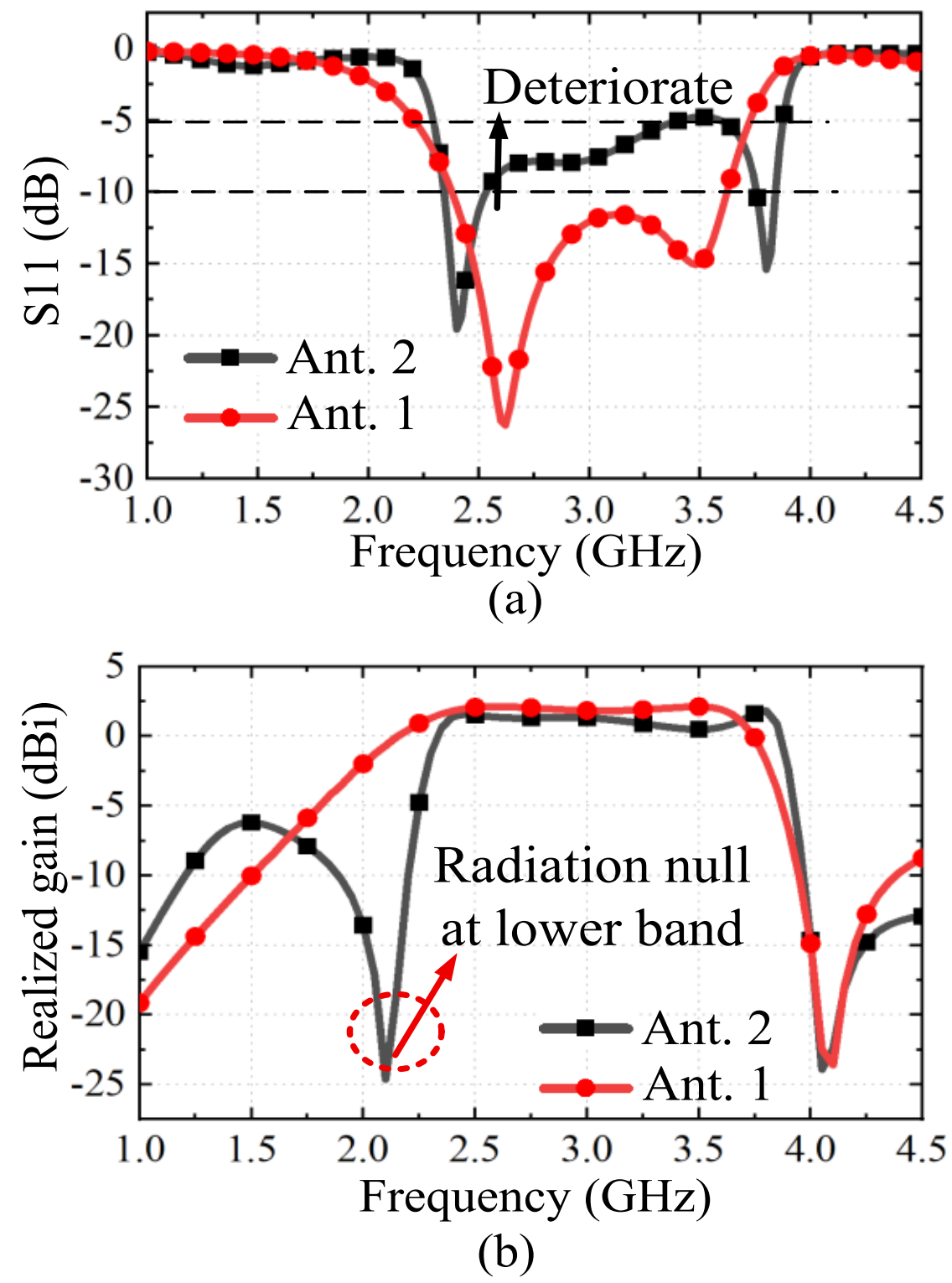

Fig. 7. S11 and realized gain results from Ant. 1 to Ant. 2.

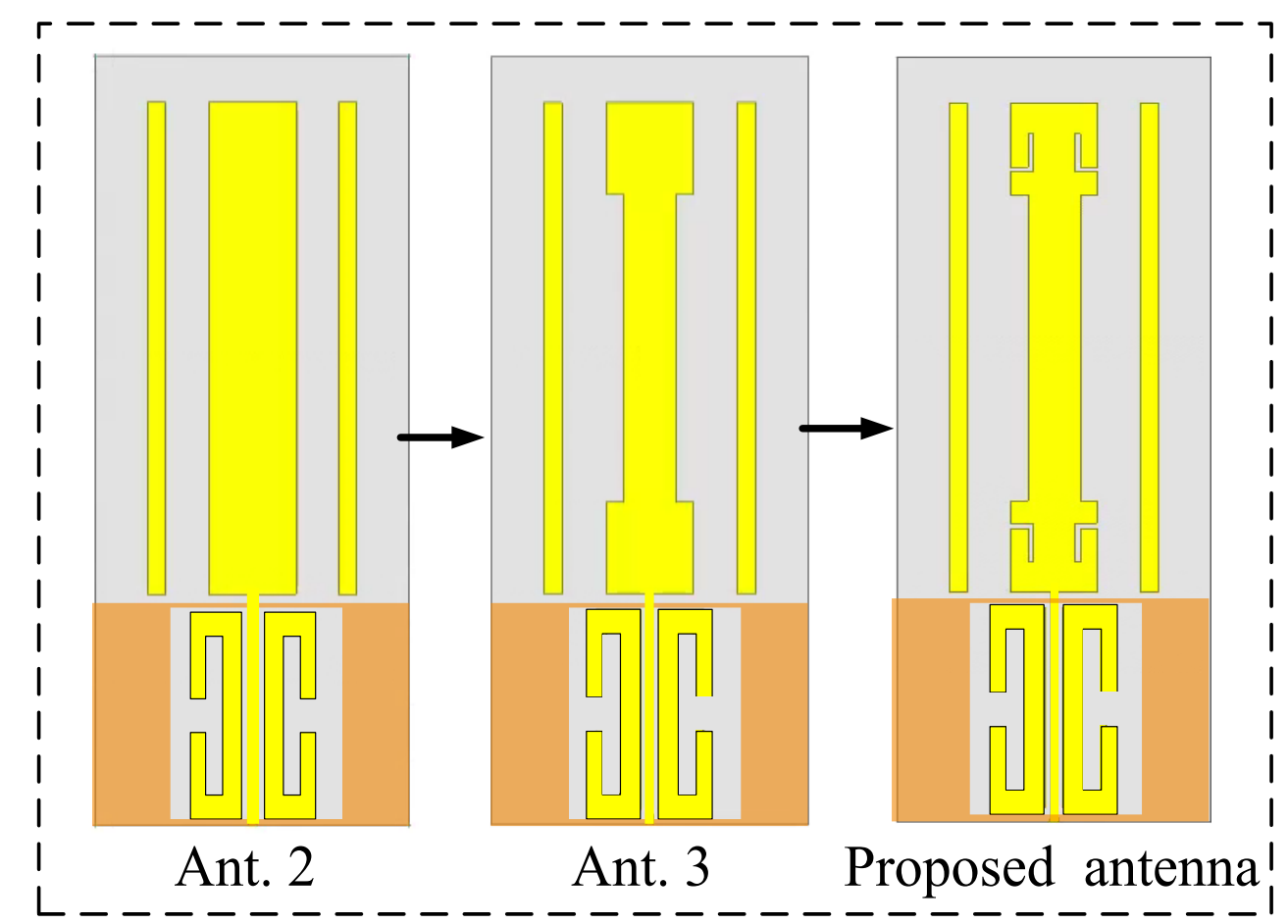

Fig. 8. Design procedure of the slotting configurations applied to Ant. 2 for a wide -10 dB impedance bandwidth.

To improve the insertion loss performance of the antenna over a wide bandwidth, slotting designs are implemented on the driven monopole of Ant. 2. Fig. 8 shows the design procedure of wideband filtering antenna achieved by adding slots on the driven monopole. It is worth noting that, the slotting configurations applied to the driven monopole also adheres to the principle of symmetry in order to maintain high omnidirectionality. Firstly, the driven monopole is slotted on both sides as antenna structure evolving from Ant. 2 to Ant. 3. As shown in the Fig. 9, the S11 of Ant. 3 improved from around -5 dB to below -10 dB from 2.8 GHz to 3.7 GHz compared with Ant. 2, effectively enhancing S11 performance in the broadband. However, the S11 of Ant. 3 is still higher than -10 dB across 2.5 GHz to 2.8 GHz. To further broaden the operating band, four L-shaped slots are attached to Ant. 3. The proposed antenna demonstrates well-matched impedance (S11 < -10 dB) from 2.2 GHz to 3.5 GHz, highlighting its wideband characteristics.

The structure and specific parameters of the proposed antenna are introduced in Fig. 10. The antenna is printed on both sides of a flexible PI substrate with a dielectric constant of 3.2, a loss tangent of 0.02, and a profile of 0.1 mm. The S11 and realized gain results of the proposed antenna is depicted in the Fig. 11. It can be observed that the proposed antenna achieves a -10 dB impedance bandwidth of 45.6% (from 2.2 GHz to 3.5 GHz). Besides, in-band gain about 2.2 dBi (2–2.5 dBi) is achieved. Furthermore, the simulated out-of-band radiation suppression is more than 10 dB at the lower band and 15 dB at the higher band, respectively. As discussed above, the proposed antenna has wide operating band and effective filtering capability with symmetric filtering structure.

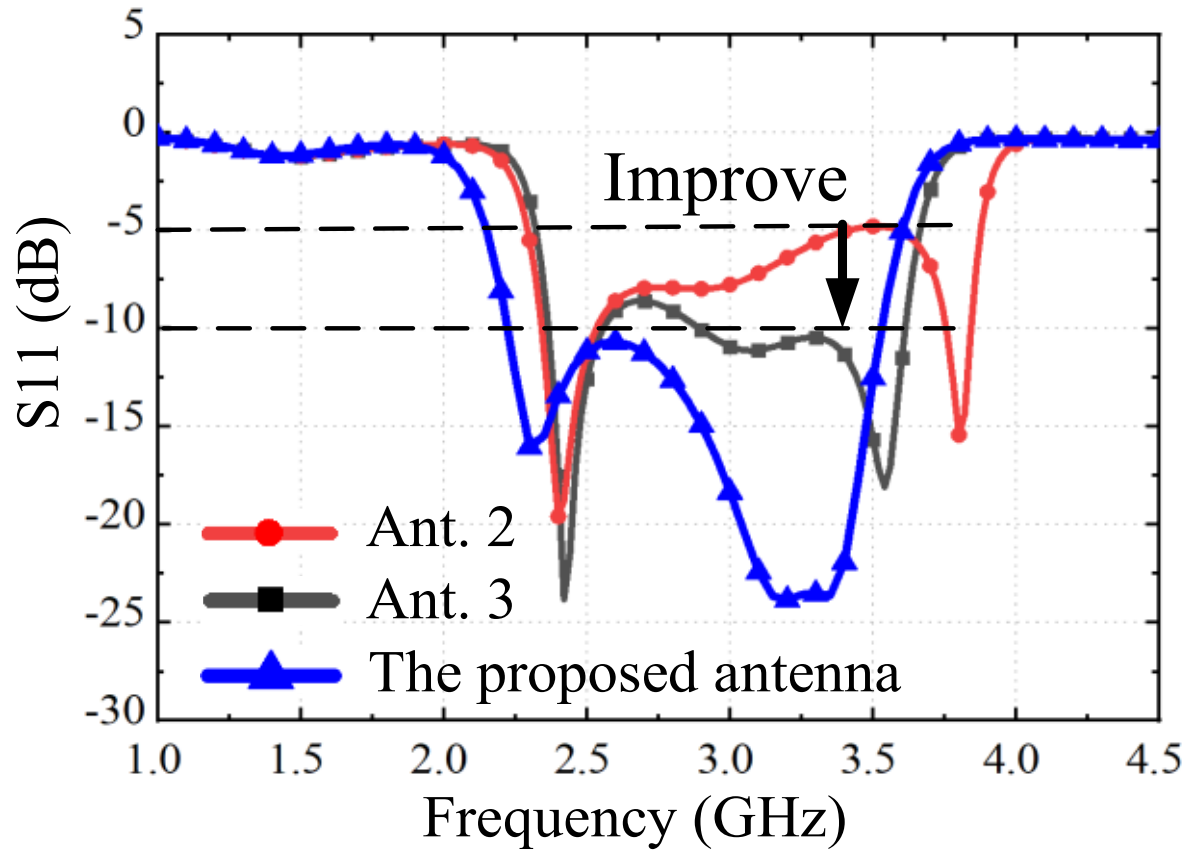

Fig. 9. Simulated S11 results for Ant. 2, Ant. 3, and the proposed antenna.

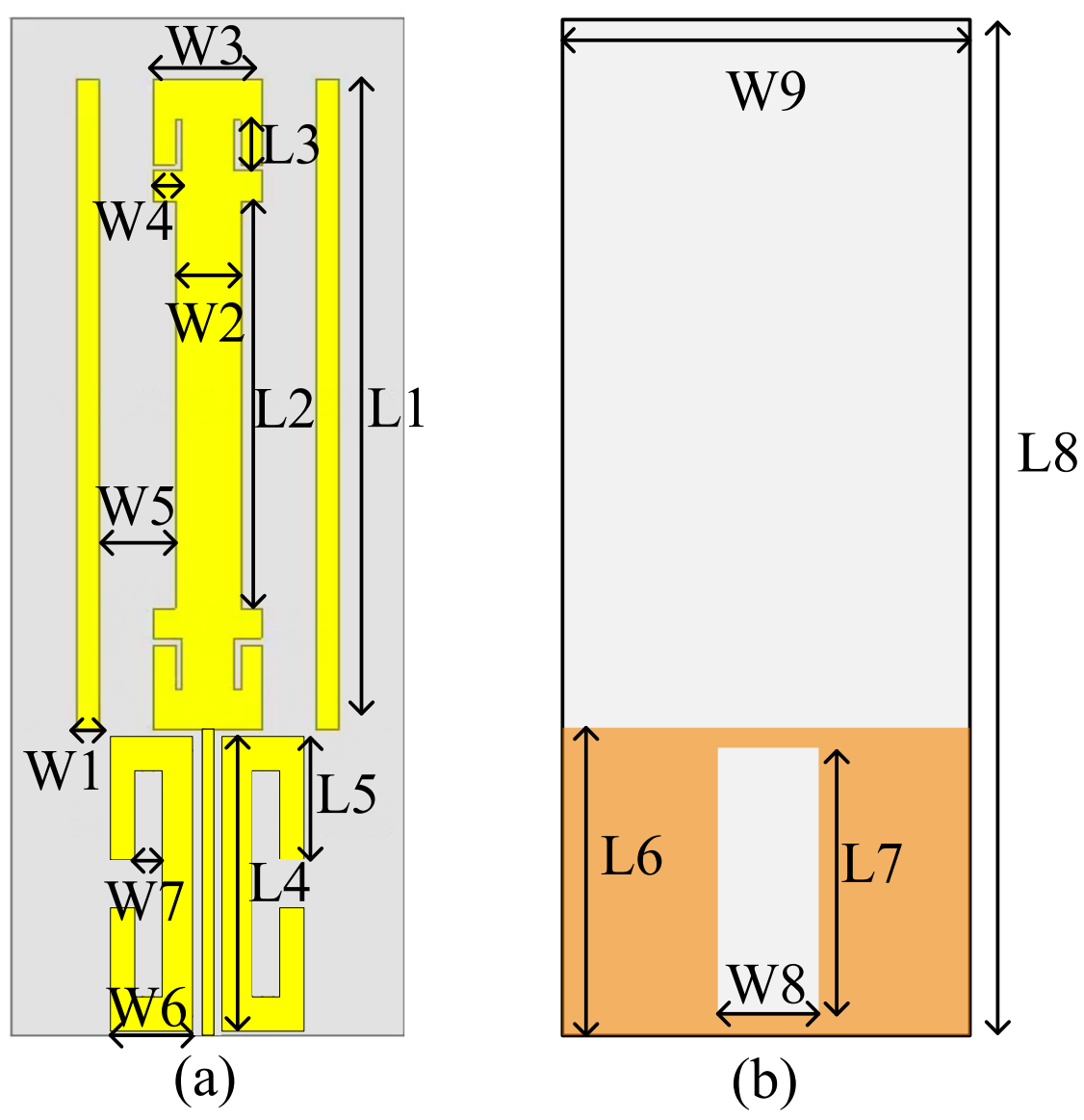

Fig. 10. Configuration of the proposed antenna. (a) top view (b) bottom view. (L1=32 mm, L2=20 mm, L3=2 mm, L4=13 mm, L5=6 mm, L6=15 mm, L7=13 mm, L8=50 mm, W1=1 mm, W2=3 mm, W3=5 mm, W4=1.3 mm, W5=3.5 mm, W6=2.5 mm, W7=1.3 mm, W8=4 mm, W9=18 mm.)

In addition to wide working band and effective out-of-band radiation suppression, the in-band normalized radiation patterns at 2.2 GHz, 2.7 GHz, 3 GHz, and 3.5 GHz are shown in the Fig. 12. The co-polarization fields have variations less than 0.2 dB, 0.2 dB, 0.3 dB, and 0.5 dB at 2.2 GHz, 2.7 GHz, 3 GHz, and 3.5 GHz in the H-plane, respectively. It can be observed that the proposed omnidirectional filtering antenna achieves high omnidirectionality with gain variation in the H-plane less than 0.5 dB across the wide operating band. As illustrated in

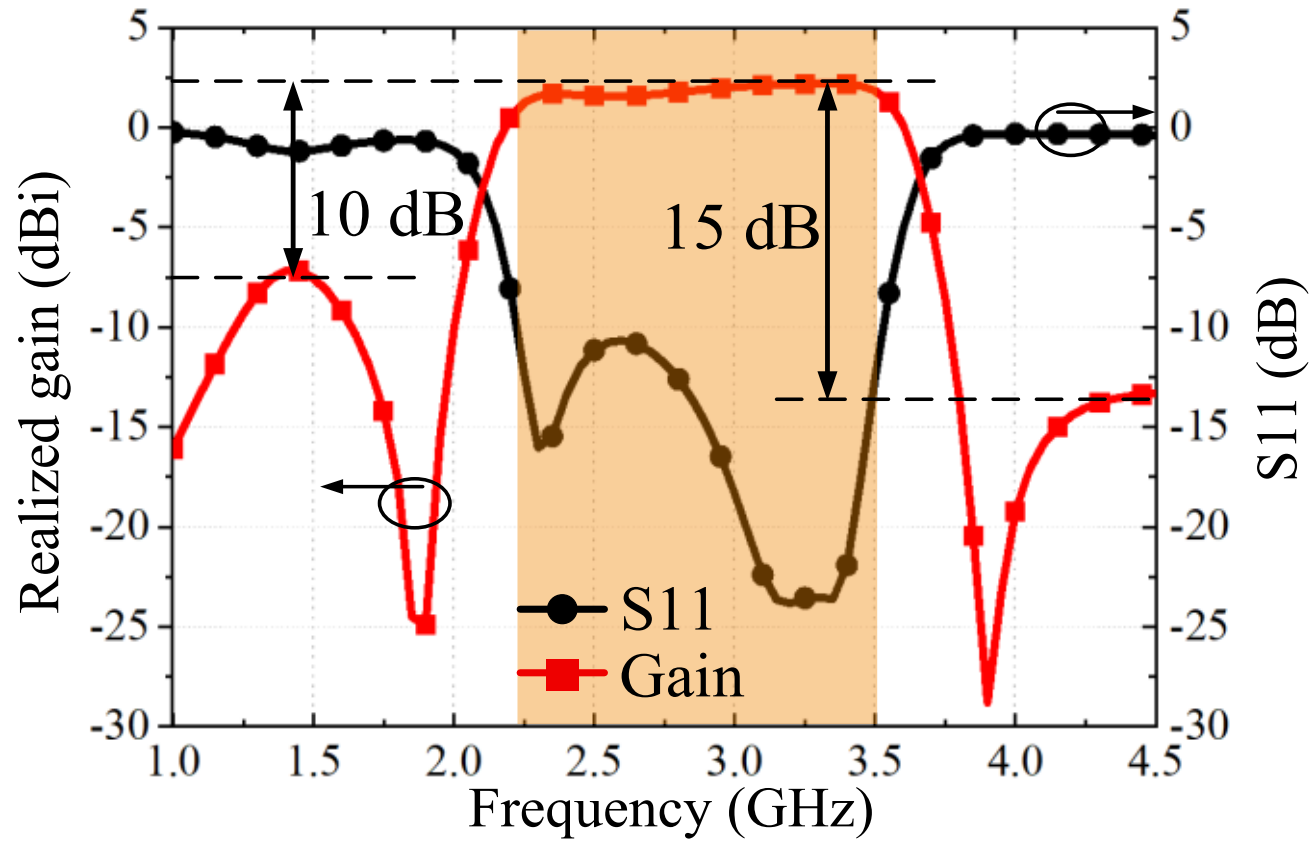

Fig. 11. Simulated S11 and realized gain results of the proposed antenna.

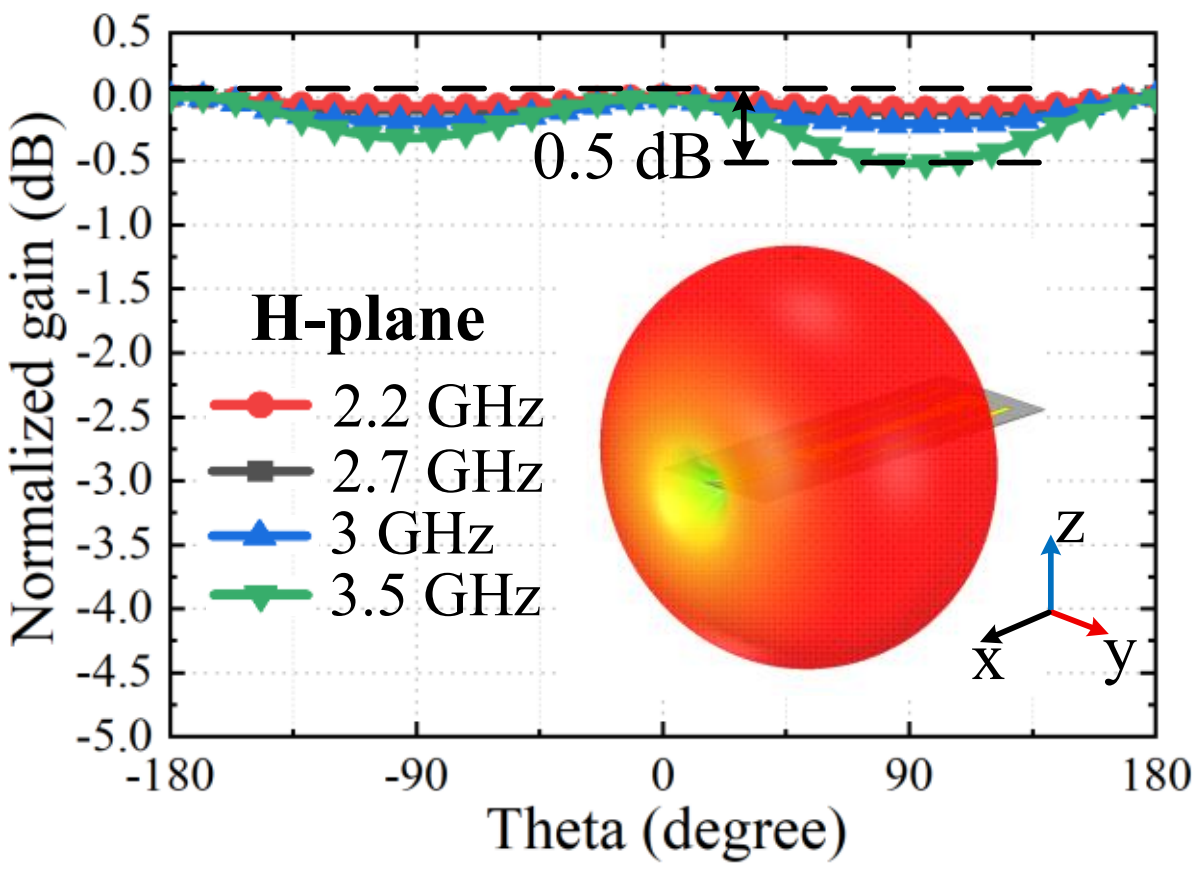

Fig. 12. Simulated in-band normalized radiation patterns in the H-plane of the proposed antenna.

Fig. 3, the in-band gain variation of the driven monopole alone is less than 0.3 dB. After integrating the symmetrical filtering structure, the gain variation of the proposed antenna becomes less than 0.5 dB, which is very close to that of the standalone driven monopole. This demonstrates that adjusting the symmetry of the parasitic elements can effectively mitigate their impact on the in-band radiation pattern of the omnidirectional filtering antenna, thereby achieving high omnidirectionality.

To further elucidate the mechanism of high in-band omnidirectionality performance realized by the symmetric filtering configuration, we add a characteristic mode analysis (CMA) for the proposed antenna, providing insights into its in-band radiation from a modal perspective. The index of modal significance (MS) is used to help us analyze the main characteristic modes, and the index of modal weighting coefficient (MWC) is used to measure the contributions of different modes to the total radiation. A larger MWC indicates a greater impact on the total radiation, while a smaller MWC suggests a minimal excitation. The MS and MWC results for the proposed antenna is provided in Fig. 13. From the MS results, it can be observed that four modes called Mode 1, Mode 2, Mode 3, and Mode 4, are primarily present within the operating band from 2.2 GHz to 3.5 GHz. Specifically, Mode 1 covers the full 2.2–3.5 GHz band. The Mode 2 mainly exists in the 2.2–3 GHz range, and Modes 3-4 mainly appear from 3 GHz to 3.5 GHz. Moreover, the MWC results reveal that from 2.2 GHz to 3 GHz, the radiation of the proposed antenna is primarily determined by Modes 1 and Mode 3. From 3 GHz to 3.5 GHz, the Modes 1 and Mode 4 become the dominant contributors. The MWC of Mode 2 is so minimal that its influence on the antenna's radiation is insignificant.

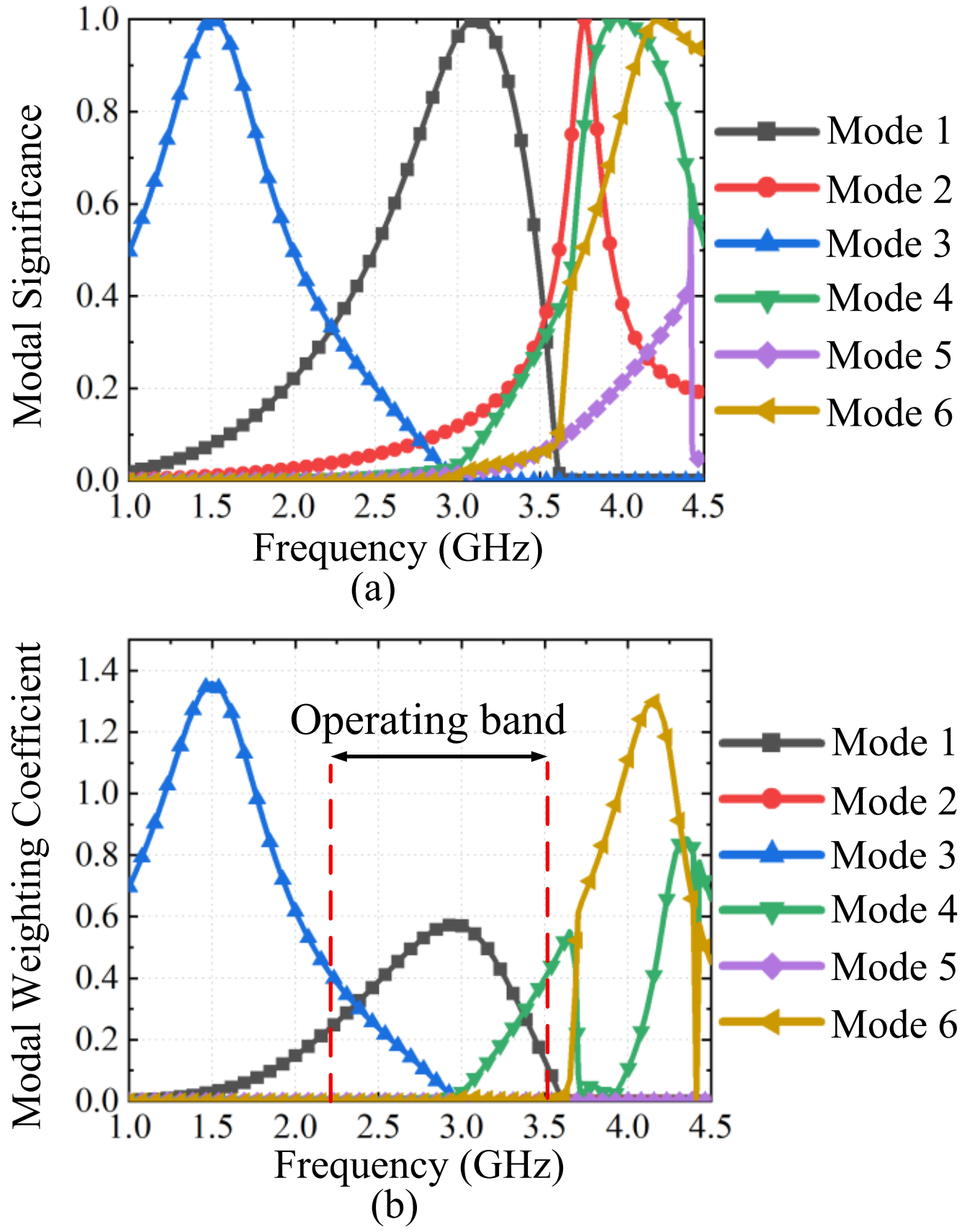

Fig. 13. Simulated modal significance and modal weighting coefficient curves for excited modes of the proposed antenna.

The radiation patterns and characteristic currents for the Mode 1, Mode 3, and Mode 4 at 2.2 GHz, 3 GHz, and 3.5 GHz are presented in the Fig. 14, Fig. 15, and Fig. 16 to explain the different radiation pattern composition in the wide operating band. These three frequency points are situated at the beginning, middle, and end of the operating band. Each frequency point differs in its mode composition, making them collectively representative of the antenna's behavior across the entire band.

Firstly, as shown in the Fig. 14, the radiation patterns and characteristic currents at 2.2 GHz are presented. The radiation patterns for Mode 1 and Mode 3 show good omnidirectional patterns, while Mode 2 exhibits a split in the middle. This can be explained by the characteristic currents distribution as discussed in the Fig. 13(b). For Mode 1 and Mode 3, the currents are concentrated on the driven monopole. Meanwhile, the currents on the symmetrically placed parasitic strips exhibit low and uniform distributions. Consequently, the radiation generated by the currents on the parasitic strips is both low in intensity and omnidirectional in the far field. Meanwhile, the symmetric coupling between the parasitic strips and the monopole also reduces the impact on the current distribution of driven monopole, thereby ensuring high omnidirectionality. In contrast, the primary currents of Mode 4 are coupled from the patch to uniformly distributed parasitic strips on both sides. Due to the symmetry of these parasitic strips, their radiation cancels out at the center, resulting in a split radiation pattern. As discussed in Fig. 13(b), the radiation of the proposed antenna at 2.2 GHz is primarily determined by Modes 1 and Mode 3. Therefore, the antenna overall exhibits good omnidirectional radiation at 2.2 GHz.

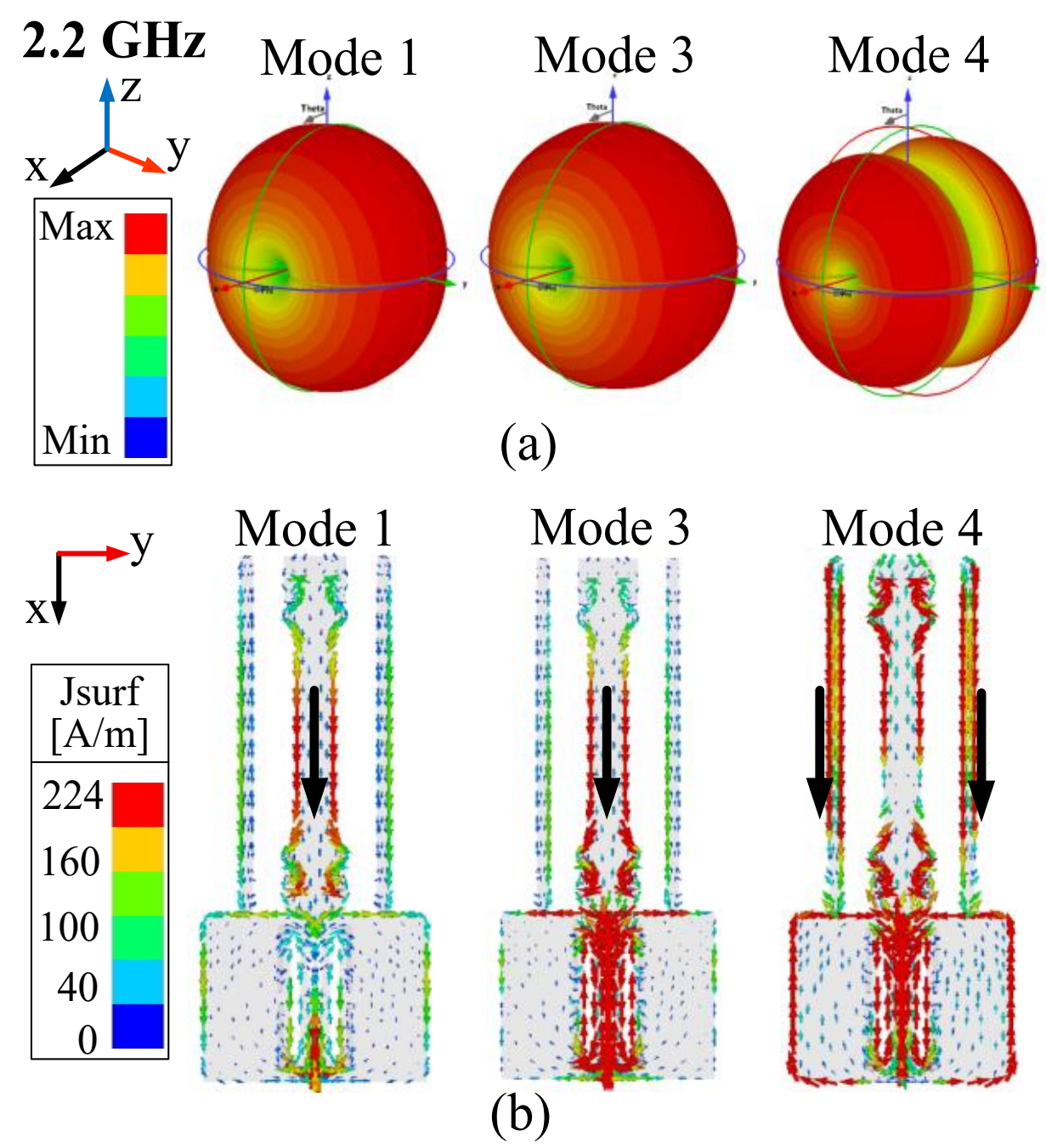

Fig. 14. Simulated in-band radiation patterns and characteristic currents for Mode 1, Mode 3, and Mode 4 at 2.2 GHz.

Besides, as shown in the Fig. 15, the radiation patterns and characteristic currents at 3 GHz are presented. As discussed in the Fig. 13(b), the Mode 1 dominates, whereas Mode 3 and 4 are insignificant. As observed from Fig. 15(a), the Mode 1 exhibits a desirable omnidirectional pattern. The symmetric parasitic strips carry weak, and uniformly distributed currents, effectively suppressing their detrimental effect on the in-band omnidirectionality. Moreover, the radiation patterns of Mode 3 begin to deteriorate and ceases to be omnidirectional. This correlates with the decline in its MWC. At frequencies above 3 GHz, the contribution of Mode 3 to the overall omnidirectional radiation becomes minimal. Therefore, the antenna overall exhibits good omnidirectional radiation.

Moreover, the radiation patterns and characteristic currents at 3.5 GHz are shown in Fig. 16. As discussed in the Fig. 13(b), the radiation of the proposed antenna at 3.5 GHz is primarily contributed by Modes 1 and Mode 4, and the MWC of Mode 3 is minimal. As shown in the Fig. 16(a), the radiation patterns of Mode 1 and Mode 4 are omnidirectional. From Fig. 16(b), it can be observed that the currents of Mode 1 are concentrated on the driven monopole, which is similar as discussed above. The currents distributions of Mode 4 exhibits some differences. At 3.5 GHz, the current is primarily concentrated on the driven monopole antenna. However, a portion of the currents is still coupled onto the symmetric parasitic strips, whose direction is opposite to the currents on the monopole. This phenomenon is attributed to the electric coupling between the parasitic strips and the driven monopole, thereby generating a radiation null at the higher band. As shown in the Fig. 13(b), the Mode 4 exhibits a pronounced MWC null at 3.8 GHz. At the higher band (3.8 GHz), the currents will be effectively coupled to the parasitic strips, which results in the generation of a radiation null and a MWC null of Mode 4. At 3.5 GHz, near the higher band edge, coupling between the parasitic strips and the driven monopole has already occurred but remains relatively weak. Owing to the adoption of a symmetrical parasitic structure, the coupled currents on the parasitic strips have identical strength, path, and direction. The energy radiated by the parasitic strips is both low in intensity and omnidirectional in the far field, which suppresses their impact on the omnidirectional radiation. Consequently, the proposed antenna exhibits omnidirectional radiation at 3.5 GHz. In summary, by introducing a fully symmetric parasitic filtering structure, the influence of radiation from coupling parasitic elements on the antenna's omnidirectional pattern is effectively suppressed, thereby achieving high omnidirectionality over a wide operating band.

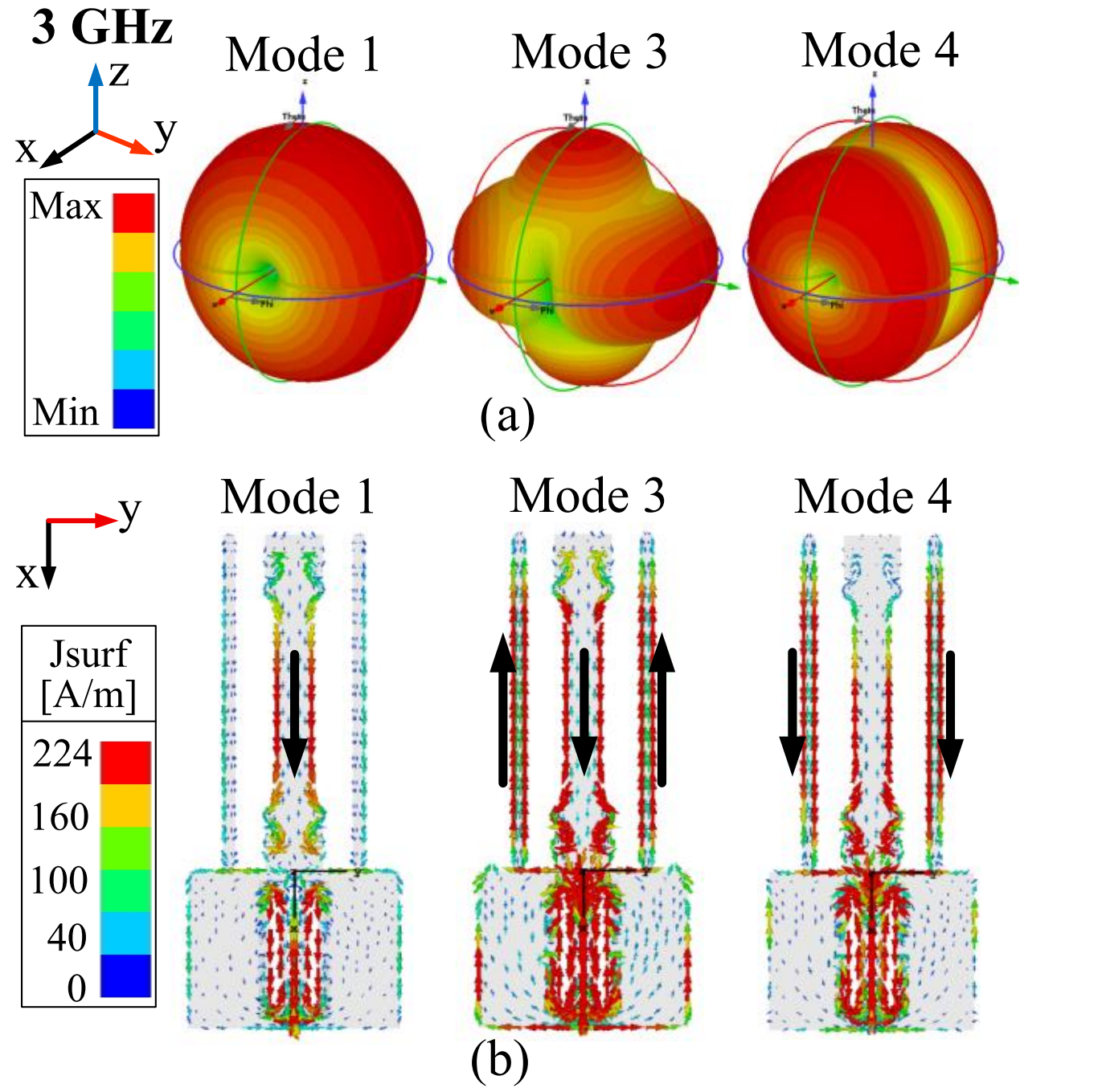

Fig. 15. Simulated in-band radiation patterns and characteristic currents for Mode 1, Mode 3, and Mode 4 at 3 GHz.

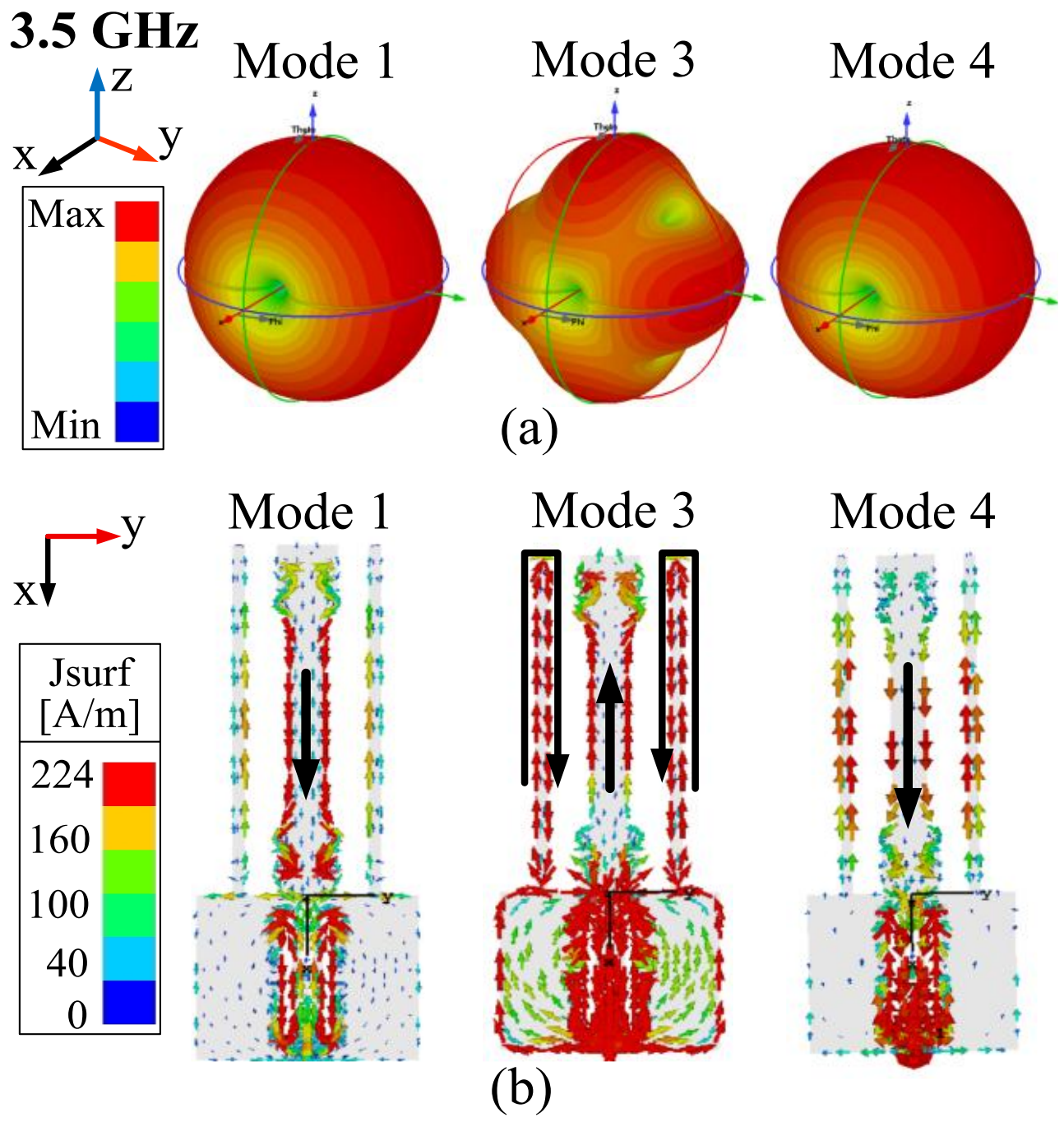

Fig. 16. Simulated in-band radiation patterns and characteristic currents for Mode 1, Mode 3, and Mode 4 at 3.5 GHz.

Besides, the in-band feeding mechanism and the filtering principle of the antenna is worth to further investigate. As the antenna structure progresses from Ant. 1 to Ant. 2 shown in Fig. 5, a pair of folded parasitic strips are added on both sides of the feed line and slots are introduced in the ground plane. This structure evolution not only generates a lower band radiation null as shown in Fig. 7, but also marks a shift in the feeding configuration from microstrip feeding to coplanar waveguide (CPW) feeding structure. The folded parasitic strips and slotted ground not only function as a filtering structure to generate a radiation null out-of-band, but also serve as a in-band driving element to support wideband CPW feeding. To better elucidate the underlying principles, we will first present the in-band feeding mechanism and the filtering principle for the lower band radiation null. Then the mechanism behind the higher band radiation null will be followed.

The principle of the in-band feeding mechanism of the proposed antenna is depicted in Fig. 17. As shown in Fig 17(a), appreciable current is observed on the driven monopole, indicating that its radiation is effectively excited. Additionally, significant current is also present on the folded parasitic strips on both sides of the feed line. Fig. 17(b) shows the magnified view of the surface currents on the slotted ground and the folded parasitic strips. On the slotted ground plane, the current is observed to flow in the -x direction. Notably, the current concentrates along the both edges of the slotted structure, indicating coupling with the folded parasitic strips above. The current on the slot is in the same direction as that on the ground plane. The outer side of the symmetric folded parasitic strip couples with the slot on the ground, inducing currents in the +x direction. Moreover, the coupling current forms a loop on the folded parasitic strips. The current loops on the symmetric parasitic strips, together with the central feed line, form a CPW feeding structure. The current on the feed line flows along +x direction, opposite to the current direction on the ground plane. Therefore, the feed line and ground work as the positive and negative poles of antenna with enabling efficient power delivery to the radiating patch. For a clearer illustration of the coupling mechanism, a flowchart of the current coupling mechanism is presented in the Fig. 17(c). The slotted ground couples with the symmetric strips, with opposite current direction. Then, the coupled current forms a loop on the folded parasitic strips, introducing a 180° phase shift. Finally, the symmetric current loops excited the central feed line with a 180° phase shift. The schematic of coupling on one side of the symmetric parasitic structure is shown in Fig. 17(d). It is noteworthy that the slot on the ground plane couples only with the outer edges of the folded parasitic strips. The coupling current on the outer sides undergoes a direction reversal on the inner side through the current loop on the folded parasitic strips, exciting the feed line with a CPW feeding structure.

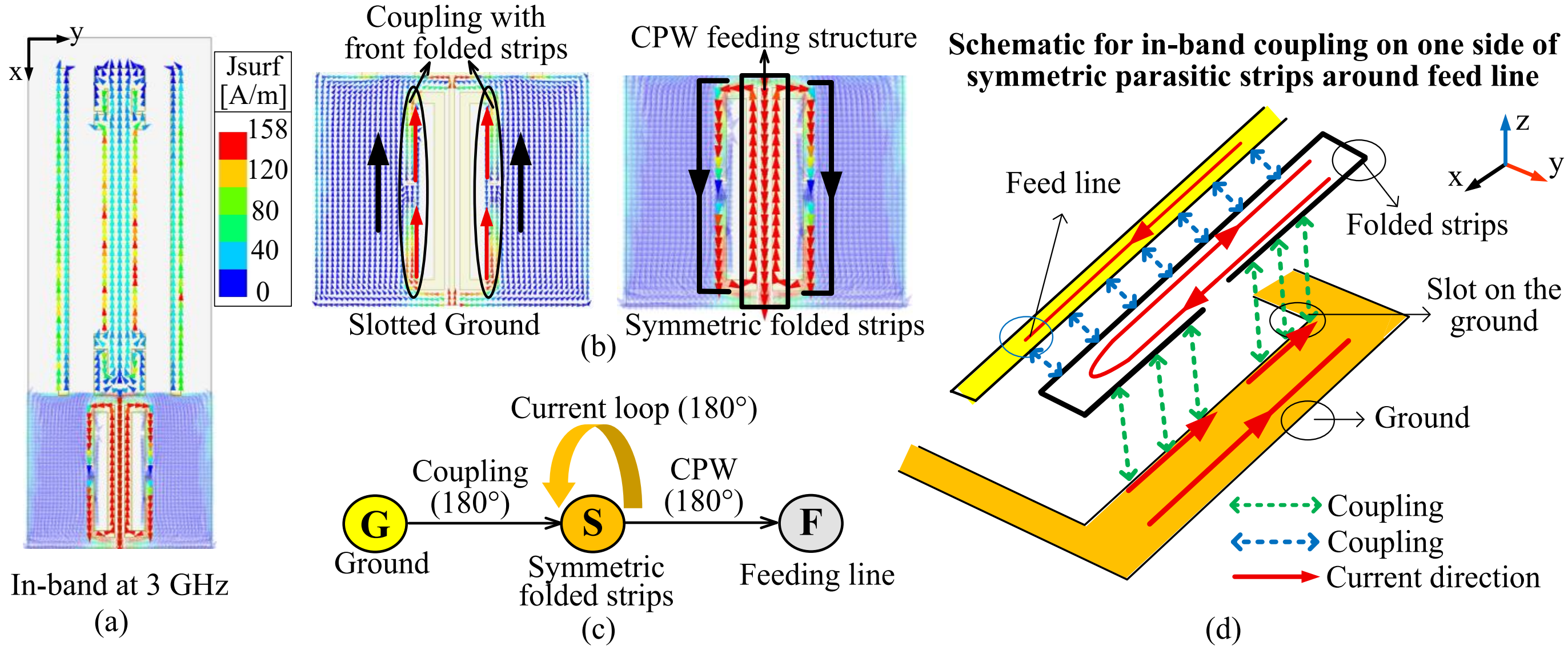

Fig. 17. Principle of the in-band feeding mechanism of the proposed antenna. (a) In-band surface current distribution. (b) Magnified view of the surface currents on the slotted ground plane (back side) and the folded parasitic strip (front side). (c) Flowchart of the current coupling mechanism. (d) Schematic of coupling on one side of the symmetric parasitic structure.

The folded parasitic strips and slotted ground not only function as in-band driving element to support wideband CPW feeding structure, but also serve as a filtering structure to generate a radiation null at the lower band. The principle of the out-of-band filtering mechanism at the lower radiation null (1.8 GHz) is shown in the Fig. 18. In contrast to the in-band current distribution shown in Fig. 17(a), the current intensity on the driven monopole at the lower band is significantly lower as shown in the Fig. 18(a). This indicates that the input power is not efficiently delivered to the monopole for radiation, thereby achieving a null. As shown in the Fig. 18(b), the current is observed to flow in the +x direction on the ground plane. However, the current direction at the both edges of slot is opposite to the ground plane, which flows along the -x direction. The current on the both edges of slot couples with the front folded strips with a 180° phase shift. Therefore, the coupling currents on the outer sides of the folded strips flowing along the +x direction, which is consistent to that on the ground plane. After passing through the current loop, the current excited by the CPW structure flows along +x direction on the feed line, which is aligned with the direction of the ground plane current. Flowchart of the current coupling mechanism at lower band radiation null is presented in the Fig. 18(c). Due to the in-phase coupling between the folded parasitic strip and ground, the phase of the feed line is identical to the ground plane. As depicted in the Fig. 18(d), the currents on the slot flow opposite to that on the ground plane. Besides, the currents on the slots couple with the outer edges of the folded parasitic strips with 180° phase shift. Therefore, the currents on the feed line are in the same direction as that on the ground plane, forming a short circuit configuration in the input port. In summary, this filtering mechanism relies on the coupling between the folded parasitic strips and the slotted ground. By modifying the current direction on the ground slot and parasitic strips at resonance, phase control between the feed line and ground is achieved. By adjusting the length of folded parasitic strips on the outer edges, which is denoted as L5 in the Fig. 10, the in-phase coupling frequency of the symmetric folded strips can be controlled. Therefore, the frequency of the lower band null can be shifted by adjusting L5. As shown in the Fig. 19, the radiation null at the lower band shifts from 1.9 GHz to 2.3 GHz when L5 decreases from 5.5 to 4 mm. It can be seen that frequencies of two radiation nulls can be effectively selected by changing the lengths of folded parasitic strips.

Following the explanation of the principle behind the lower band radiation null, the filtering mechanism for the higher band is discussed below. The surface current distributions of the proposed antenna at the higher band radiation null are shown in the Fig. 20. The surface currents distributed on the driven monopole and the symmetric linear parasitic strips have similar amplitude with anti-phase at higher radiation null (3.8 GHz), which means that the radiation that comes from the driven monopole and symmetric strips will cancel with each other in the far-field to generate a radiation null. Moreover, as shown in the Fig. 13, Mode 4 serves as the dominant radiation mode in the operating band. In the higher band, the currents are coupled to the parasitic strips. As a result, effective suppression of out-of-band radiation is evidenced by the MWC null observed for Mode 4 at higher band, indicating that its contribution to the radiation is suppressed. Besides, by adjusting the length of symmetric linear strips, which is denoted as L1 in the Fig. 10, the frequency of the higher band radiation null can be controlled. As shown in Fig. 21, the radiation null at the higher band moves from 4.3 GHz to 3.9 GHz with L1 increasing from 30 to 32 mm. From the results in Fig. 19 and Fig. 21, it can be seen that frequencies of two radiation nulls can be significantly selected by changing the lengths of corresponding parasitic strips. Besides, changing the frequency of one null exhibit minimal impact on the other. The frequencies of the two radiation nulls can be adjusted independently by varying the length of their corresponding symmetric parasitic strips.

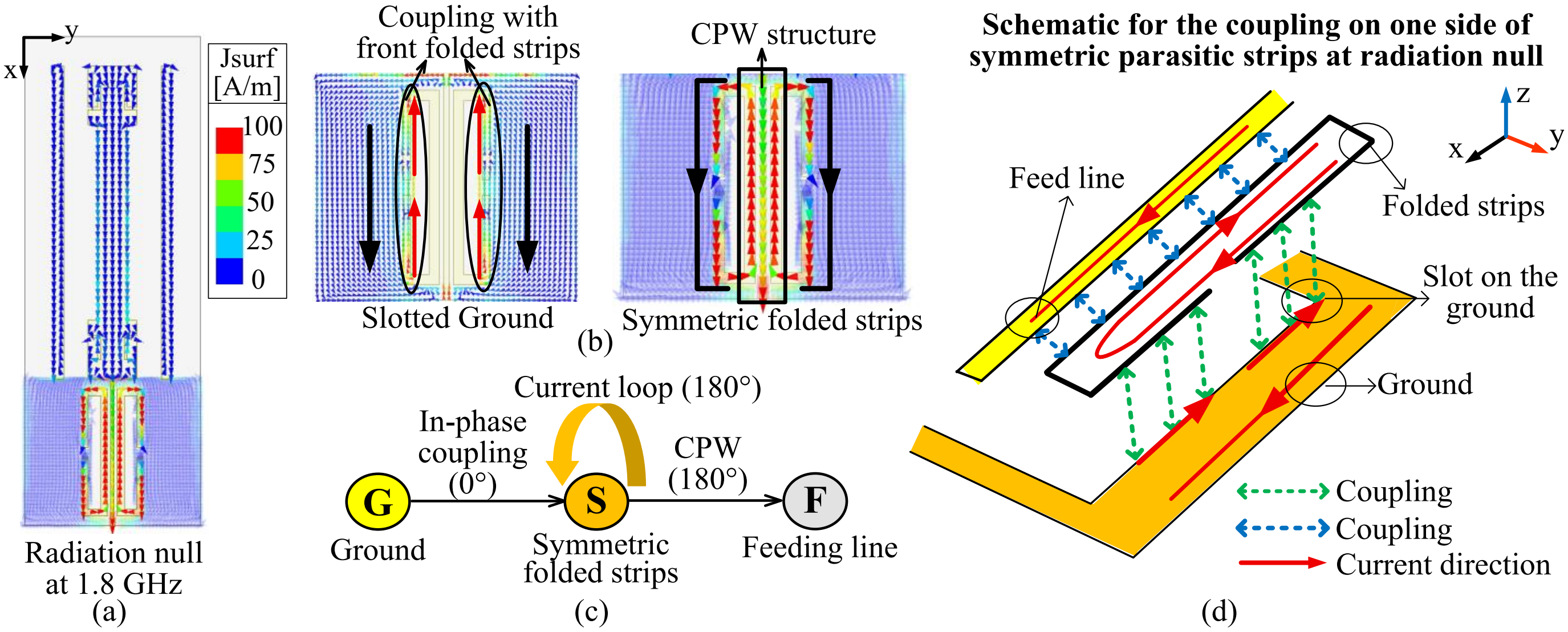

Fig. 18. Principle of the out-of-band filtering mechanism at the lower radiation null (1.8 GHz) of the proposed antenna. (a) Surface current distribution. (b) Magnified view of the surface currents on the slotted ground plane (back side) and the folded parasitic strip (front side). (c) Flowchart of the current coupling mechanism. (d) Schematic of coupling on one side of the symmetric parasitic structure.

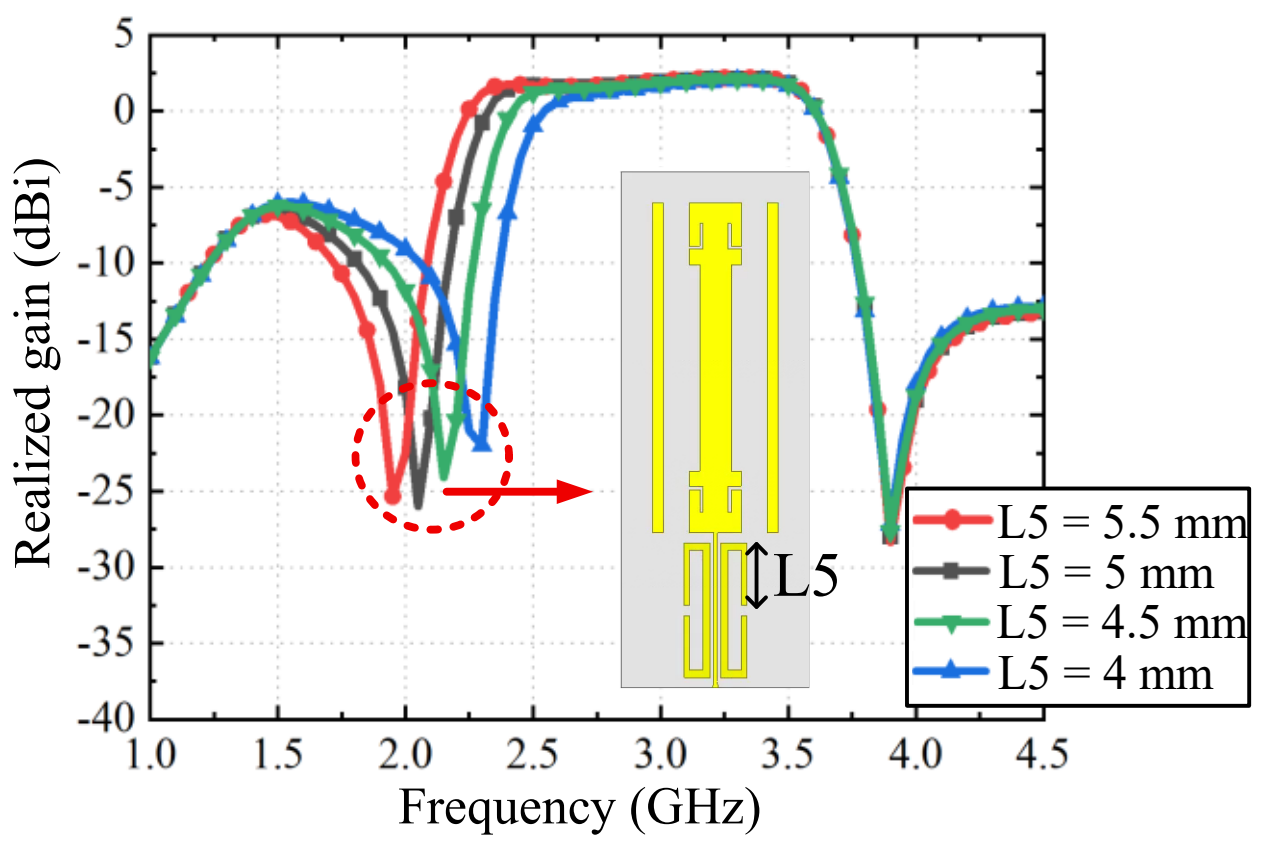

Fig. 19. Frequency adjustment of the lower band radiation null with different lengths of folded parasitic strips (L5).

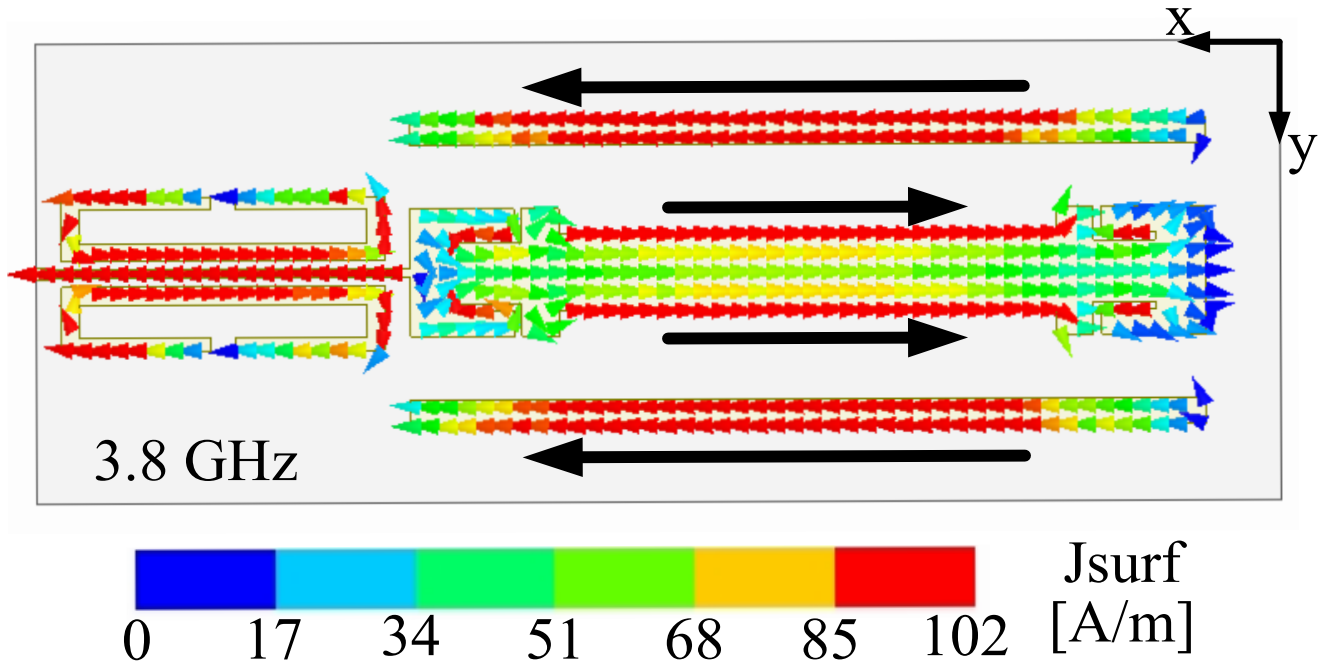

Fig. 20. Surface current distributions of the proposed antenna at the higher band radiation null (3.8 GHz).

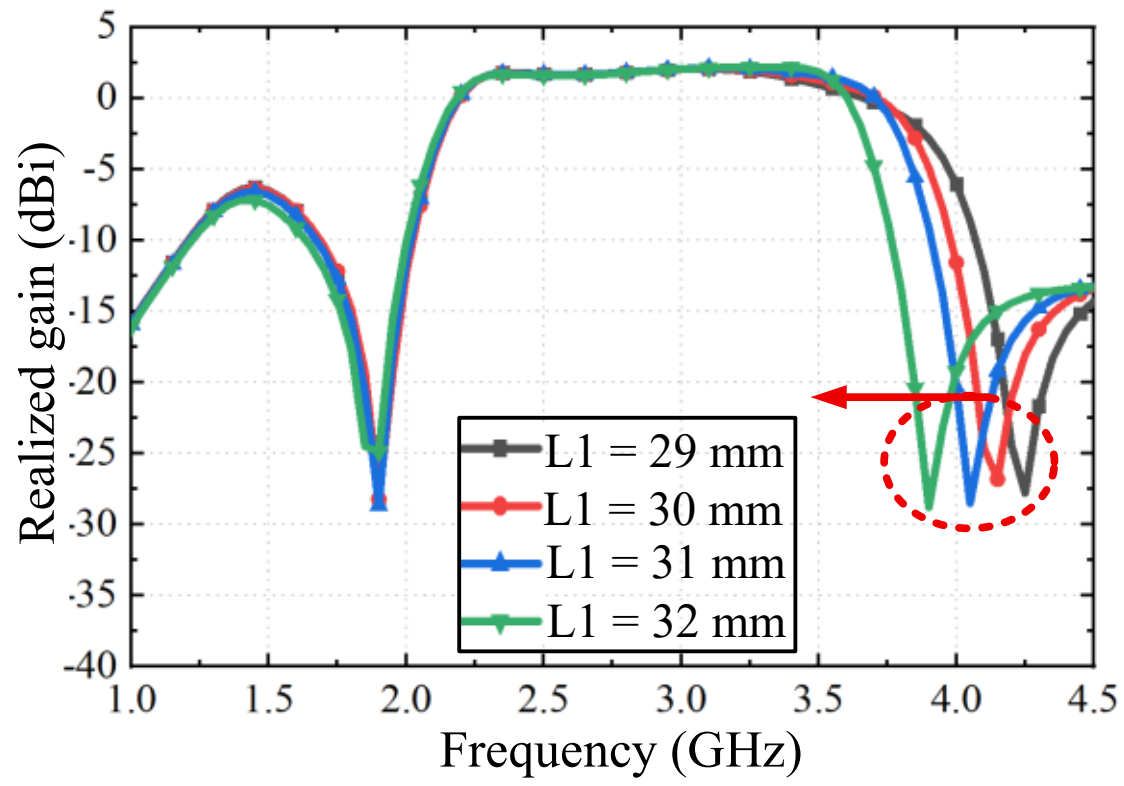

Fig. 21. Frequency adjustment of the higher band radiation null with different lengths of linear parasitic strips (L1).

### *C. Simulations of Antenna Performance After Bending*

To evaluate the effect of bending on the antenna's radiation characteristics, the proposed antenna is conformally wrapped around cylindrical surfaces with different curvature radii in the H-plane as shown in Fig. 22. The simulated S11 results are presented in Fig. 22(a). Across different bending radii, the operating band of the antenna does not exhibit significant frequency shift or performance degradation. The bent antenna demonstrates a stable -10 dB impedance bandwidth of 45.6% (from 2.2 GHz to 3.5 GHz), which is consistent with unbending conditions. Furthermore, the realized gain results of the bent antenna are depicted in Fig. 22(b). It can be observed the realized gain of the bent antenna keeps stable compared with unbending condition. The out-of-band radiation suppression at the higher and lower band are more than 15 dB and 10 dB, respectively. The realized gains from 2.1 dBi to 2.5 dBi are achieved in the wide operating band, which fluctuates slightly after bending.

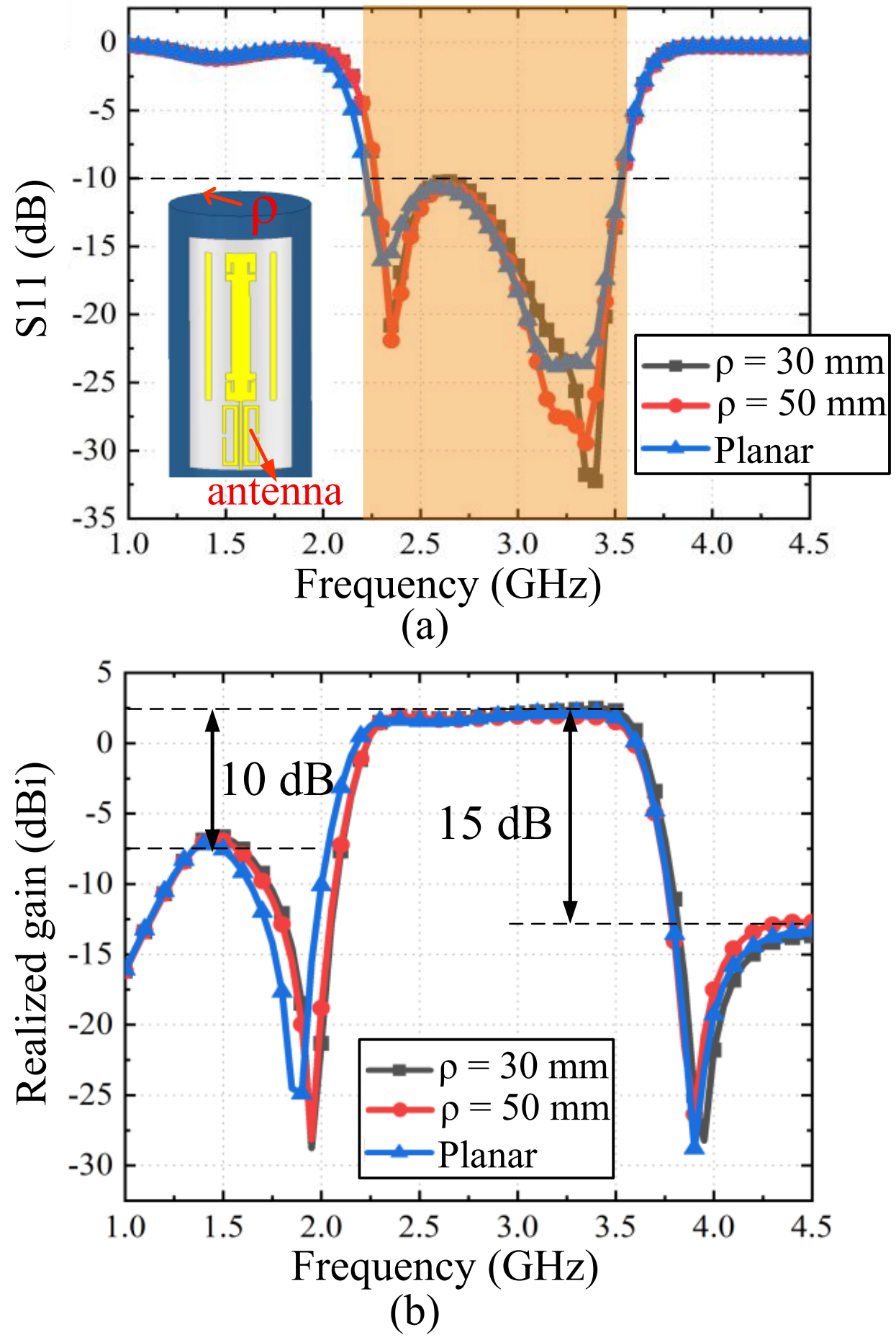

Fig. 22. S11 and Realized gain results with different bending radii. (a) S11. (b) Realized gain.

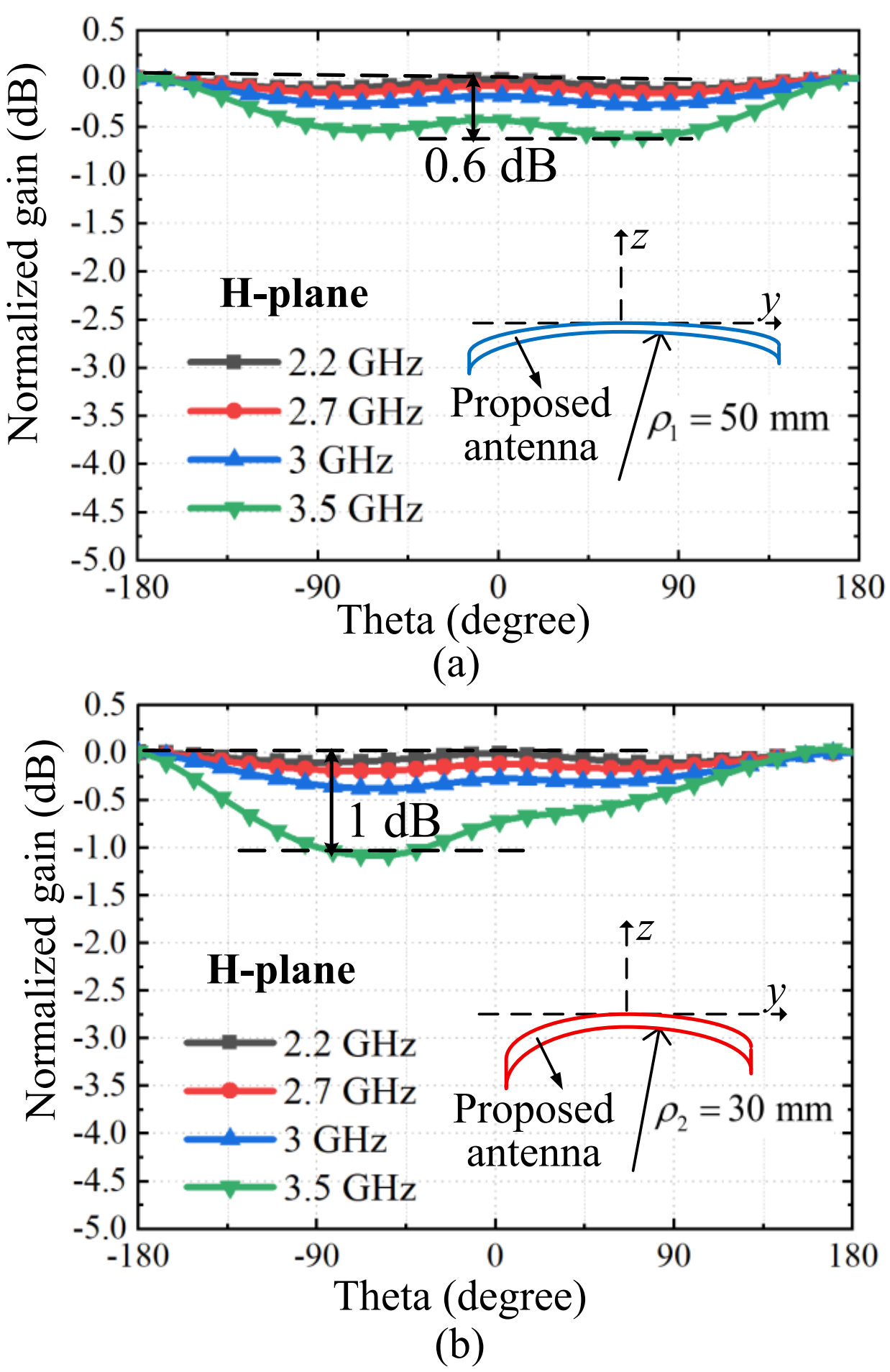

Fig. 23. In-band normalized radiation patterns of the proposed antenna under different bending radii ρ in the H-plane.

The in-band radiation patterns of the proposed antenna under different bending radii in H and E plane are shown in Fig. 23. When the ρ = 50 mm, the co-polarization fields of the proposed antenna have corresponding variations less than 0.2 dB, 0.2 dB, 0.3 dB, and 0.6 dB at 2.2 GHz, 2.7 GHz, 3 GHz, and 3.5 GHz in the H-plane. Moreover, when the bending curvature increases from ρ = 50 mm to ρ = 30mm, the co-polarization fields have variations less than 0.2 dB, 0.2 dB, 0.4 dB, and 1 dB at 2.2 GHz, 2.7 GHz, 3 GHz, and 3.5 GHz in the H-plane, respectively. As the curvature further increases, the omnidirectionality of the bent antenna remains stable around 2.2 GHz, 2.7 GHz, and 3 GHz. Moreover, a slight degradation in omnidirectionality is observed at 3.5 GHz from 0.6 dB to 1 dB. The proposed antenna exhibits good omnidirectionality with co-polarization fields variations less than 1 dB in the wideband even under different bending curvature. Through comprehensive simulation and analysis, the proposed antenna design achieves wideband flexible omnidirectional filtering capability with high omnidirectionality and stable performance after bending.

## III. Antenna Measurement And Discussion

### A. Antenna Measurement

The proposed antenna is fabricated, and its prototype is shown in Fig. 24(a). Due to the compact size of the antenna, an SMP connector is adopted. Besides, to precisely determine the antenna's curvature, a 3D printed foam cylinder is employed as depicted in Fig. 24(b). The proposed flexible antenna is conformally attached to the cylinder surface, achieving a bending radius of 30 mm. Both the flat and bent antennas are measured in a spherical anechoic chamber as shown in Fig. 24(c). The measured and simulated S11 results are shown in Fig. 25. It can be observed that the -10 dB impedance bandwidth of the proposed antenna is 45.6% (from 2.2 GHz to 3.5 GHz). Besides, the measured operating band exhibits slight variation in both flat and bent states.

The measured and simulated realized gain results are presented in Fig. 26. The proposed antenna offers a peak gain about 2.5 dBi in the operating band. Besides, the flat antenna can achieve an out-of-band radiation suppression more than 12.5 dB in the lower band and 15 dB in the higher band, respectively. When the bending radius increases to 30 mm, the peak in-band gain of the proposed antenna remains almost unchanged. Moreover, the out-of-band radiation suppression of the bent antenna is more than 11 dB in the lower band and 15 dB in the higher band, which keeps stable compared with unbending results. Fig. 27 depicts the measured in-band gain variations in the H-plane of the proposed antenna at 2.2 GHz, 3 GHz, and 3.5 GHz in the both flat and bending states. The measured in-band co-polarization fields of the flat antenna have variations less than 0.8 dB in the H-plane. There is a 0.3 dB difference between the experimental and simulated data, which can be attributed to environmental factors and manufacturing tolerances during measurement. Besides, the in-band gain variation in the H-plane is less than 1 dB with ρ = 30 mm. The measured cross-polarization levels for both flat and bent antenna are more than −15 dB in the E-plane and H-plane as shown in the Fig. 28 and Fig. 29. The levels of the simulated cross-polarization in the E-plane and the H-plane are all lower than −25 dB, which is not shown here. The measured cross polarization level is higher than the simulated results, which is attributed to measurement errors in the anechoic chamber. From the above measurement results, it can be seen that the proposed filtering antenna exhibits high in-band omnidirectionality across a wide operating bandwidth and effective out-of-band radiation suppression. Furthermore, its performance remains stable after bending.

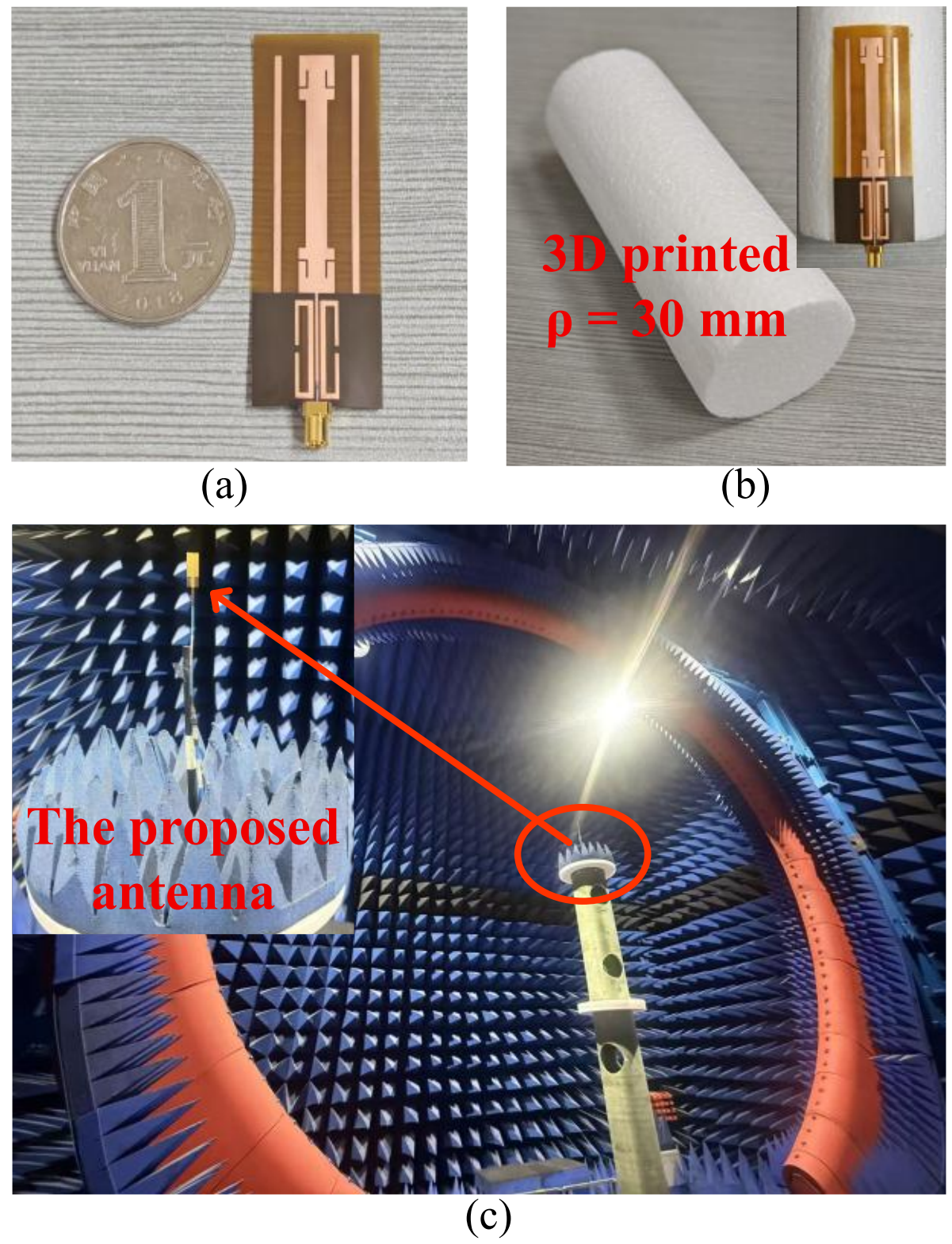

Fig. 24. Antenna prototype and its measurement setup. (a) Antenna prototype. (b) 3D printed cylinder with ρ = 30 mm. (c) Far-field testing in an anechoic chamber.

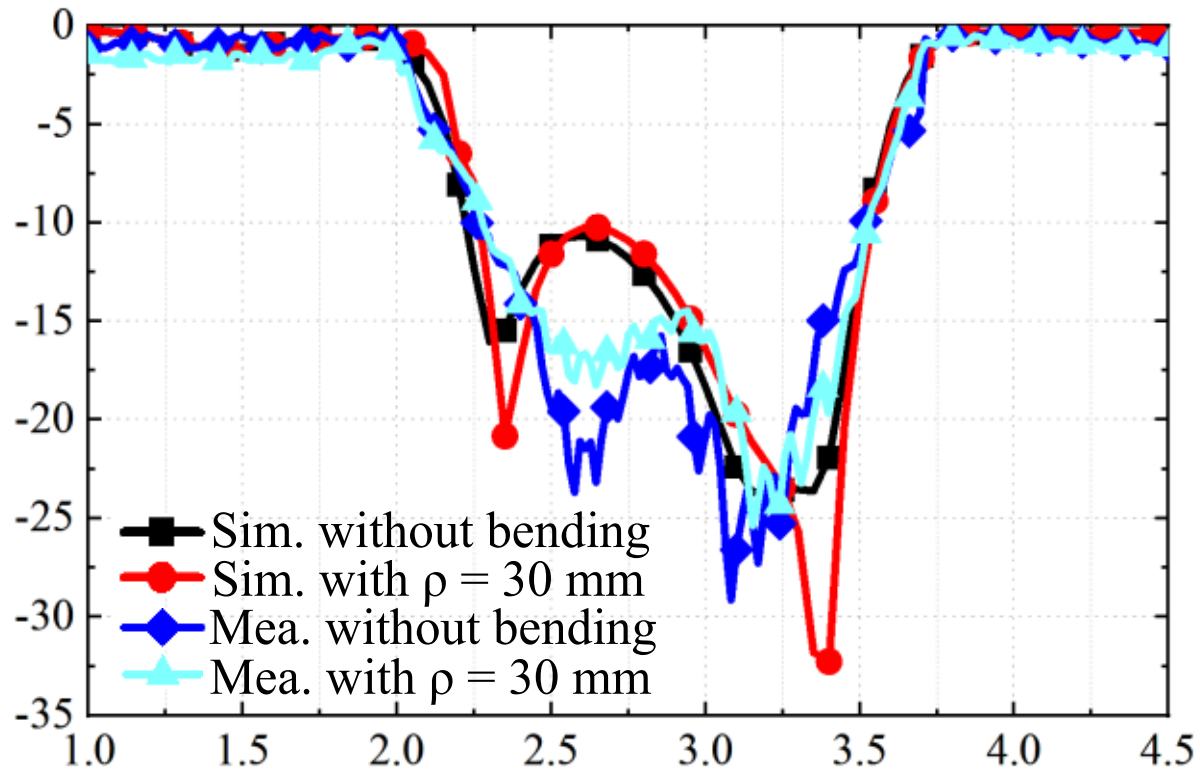

Fig. 25. Measured and simulated reflection coefficients under different bending radii.

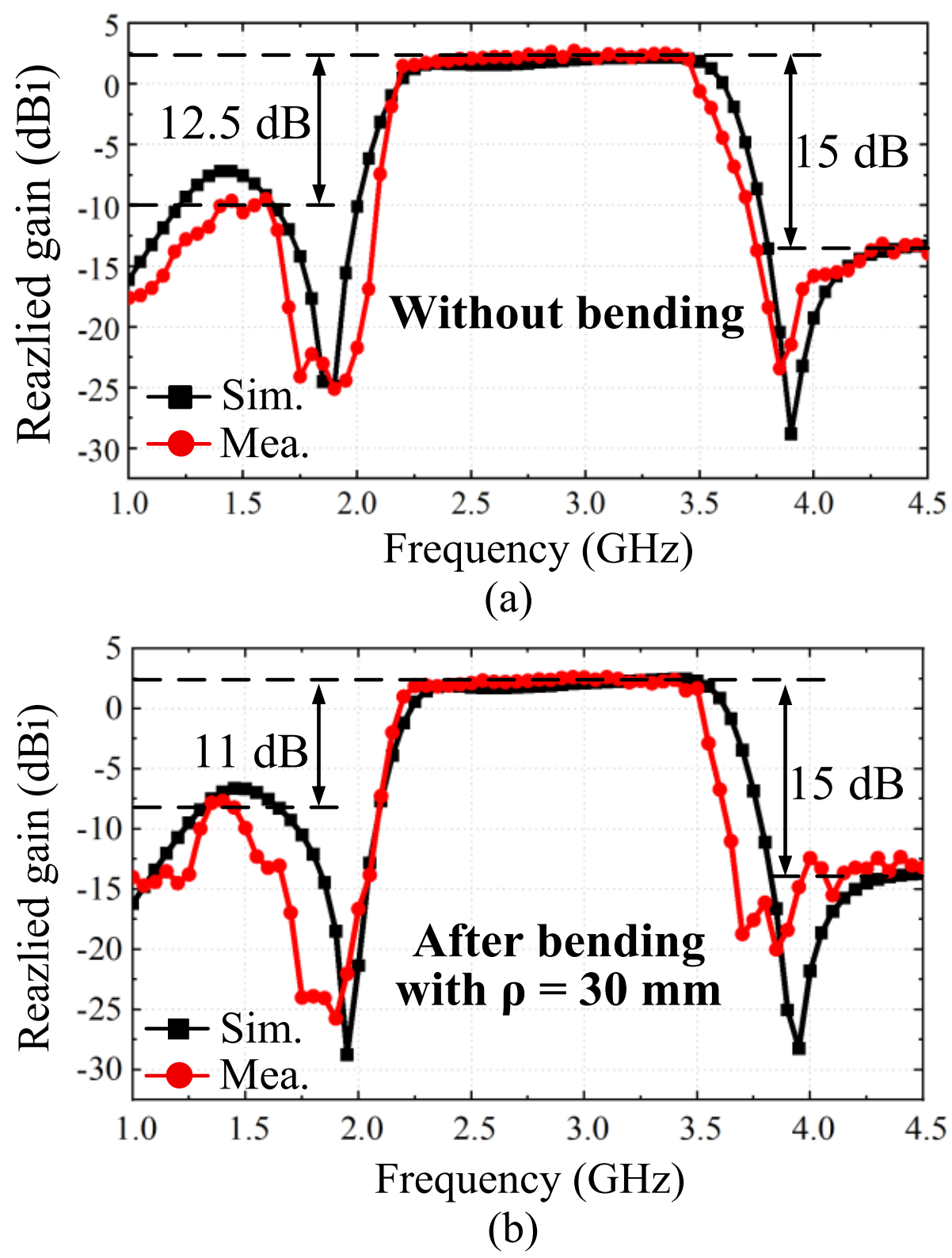

Fig. 26. Measured and simulated realized gain results. (a) Without bending. (b) After bending with ρ = 30 mm.

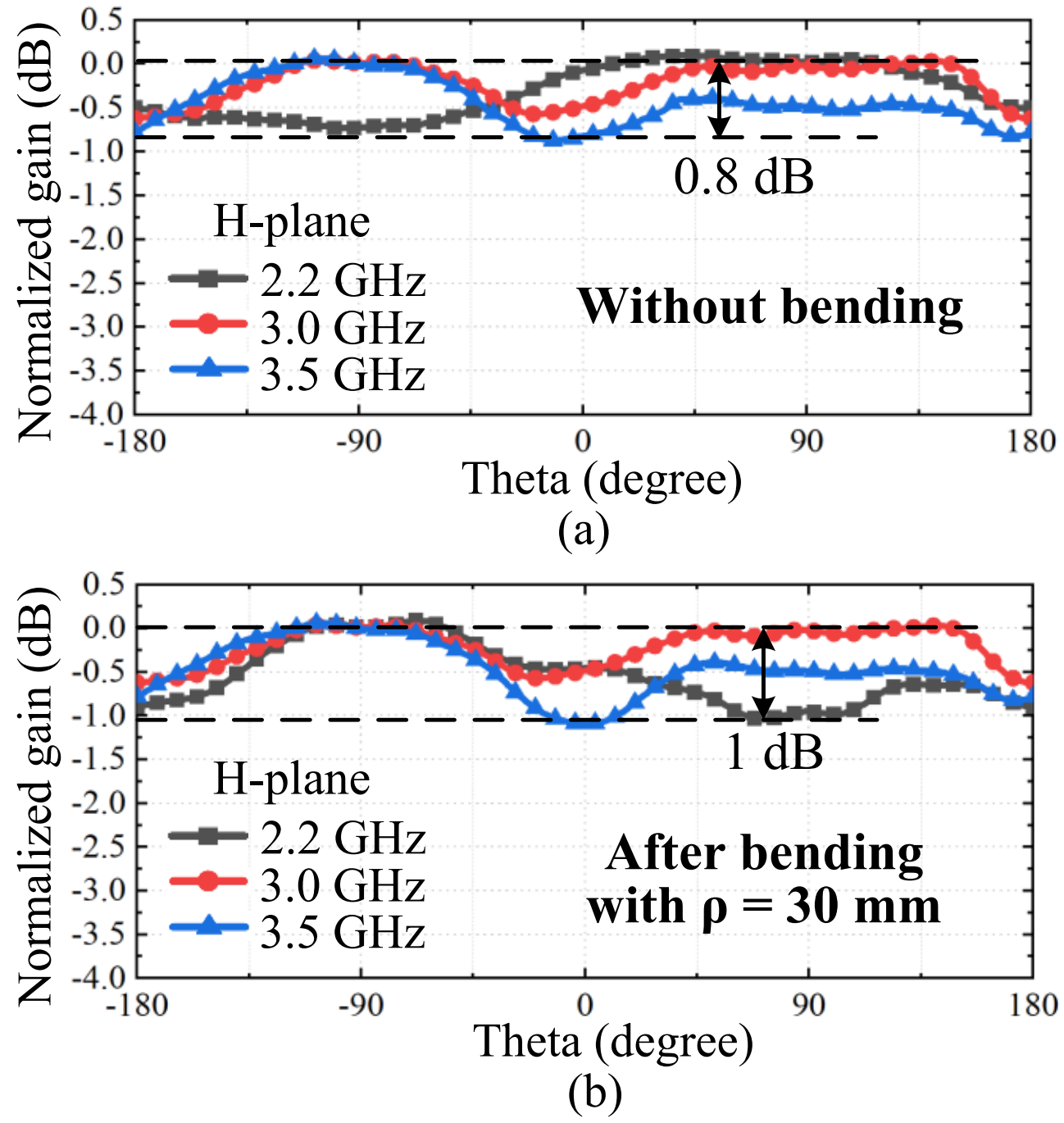

Fig. 27. Measured in-band normalized gain results of the proposed antenna in the H-plane. (a) Without bending. (b) After bending with ρ = 30 mm.

TABLE I
COMPARISON OF THE WORK WITH RELATED WORKS

| Ref. | Flexibility | Filtering Circuit | Radiation pattern | Gain Variation in the H-plane (dB) | Bandwidth | Minimum bending radius | Suppression Levels (dB) | Total Size ($\lambda_0$*$\lambda_0$) | Profile ($\lambda_0$) |
|---|---|---|---|---|---|---|---|---|---|
| [24] | No | Yes | Omni. | None | 7.9% | None | None | 0.23×0.22 | 0.01 |
| [25] | No | Yes | Omni. | <5 | 27.5% | None | >10.3 | 0.32×0.32 | 0.008 |
| [26] | No | No | Omni. | <3 | 42.5% | None | >10 | 0.4×0.006 | 0.0037 |
| [27] | No | No | Omni. | <3 | 41.2% | None | >14 | 0.44×0.44π (circular) | 0.12 |
| [28] | No | No | Omni. | <2.5 | 41.6% | None | >11 | 1.68×1.68π (circular) | 0.1 |
| [29] | No | No | Omni. | <2.2 | 14.6% | None | >14.2 | 1.61×0.13 | 0.013 |
| [31] | No | No | Omni. | <1.57 | 6.2% | None | >10 | 0.47×0.45 | 0.025 |
| [36] | Yes | No | Direct. | / | 5.7%/9.1% | 150 mm | >11 | 0.56×0.56 | 0.0004 |
| [37] | Yes | No | Direct. | / | 11.8% | 150 mm | >11 | 0.71×0.79 | 0.07 |
| **This Work** | **Yes** | **No** | **Omni.** | **<0.8(flat)/ <1(ρ=30 mm)** | **45.6%** | **30 mm** | **>12.5(flat)/ >11(ρ=30 mm)** | **0.46×0.17** | **0.0009** |

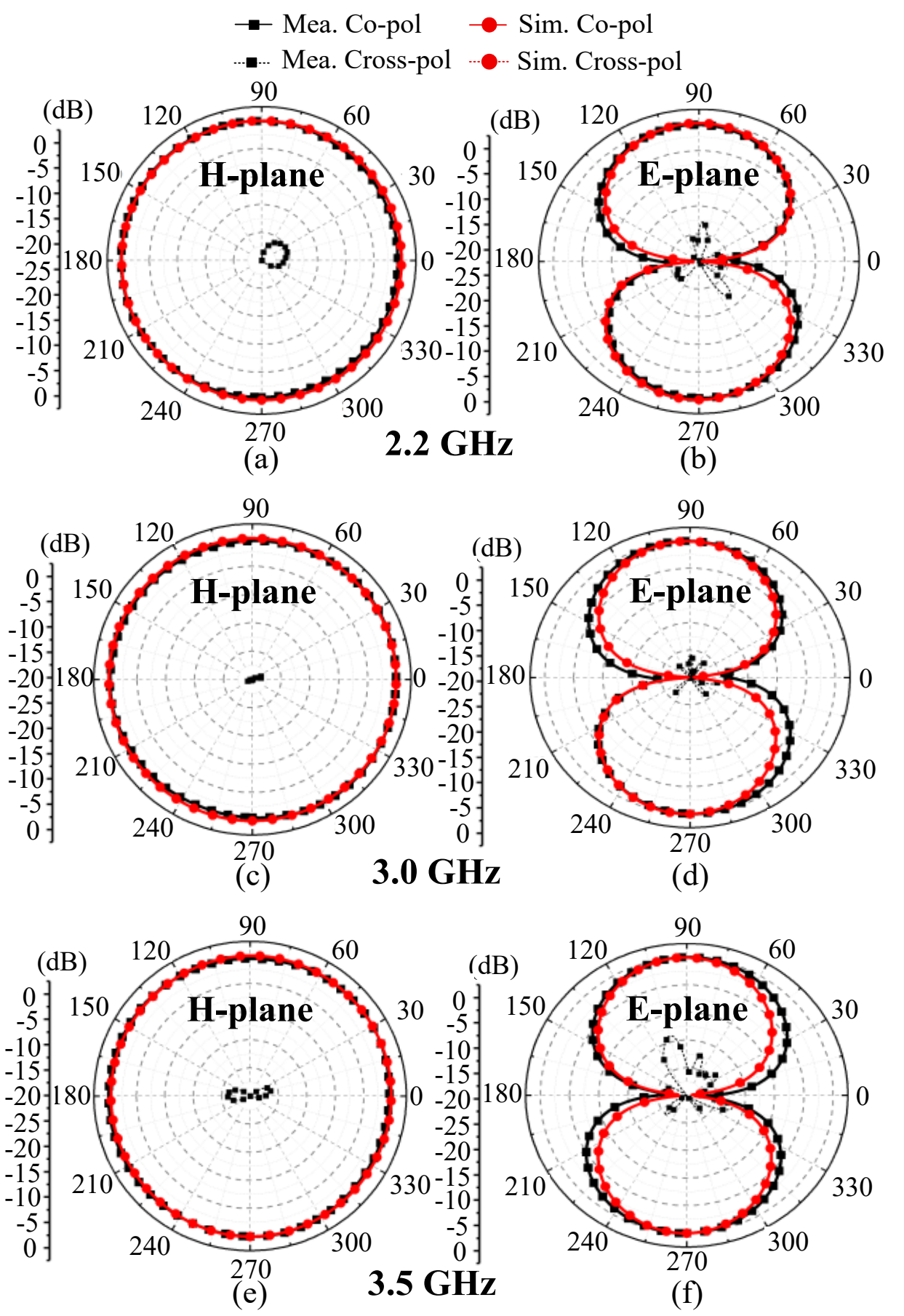

Fig. 28. Measured and simulated radiation patterns of the unbending antenna.

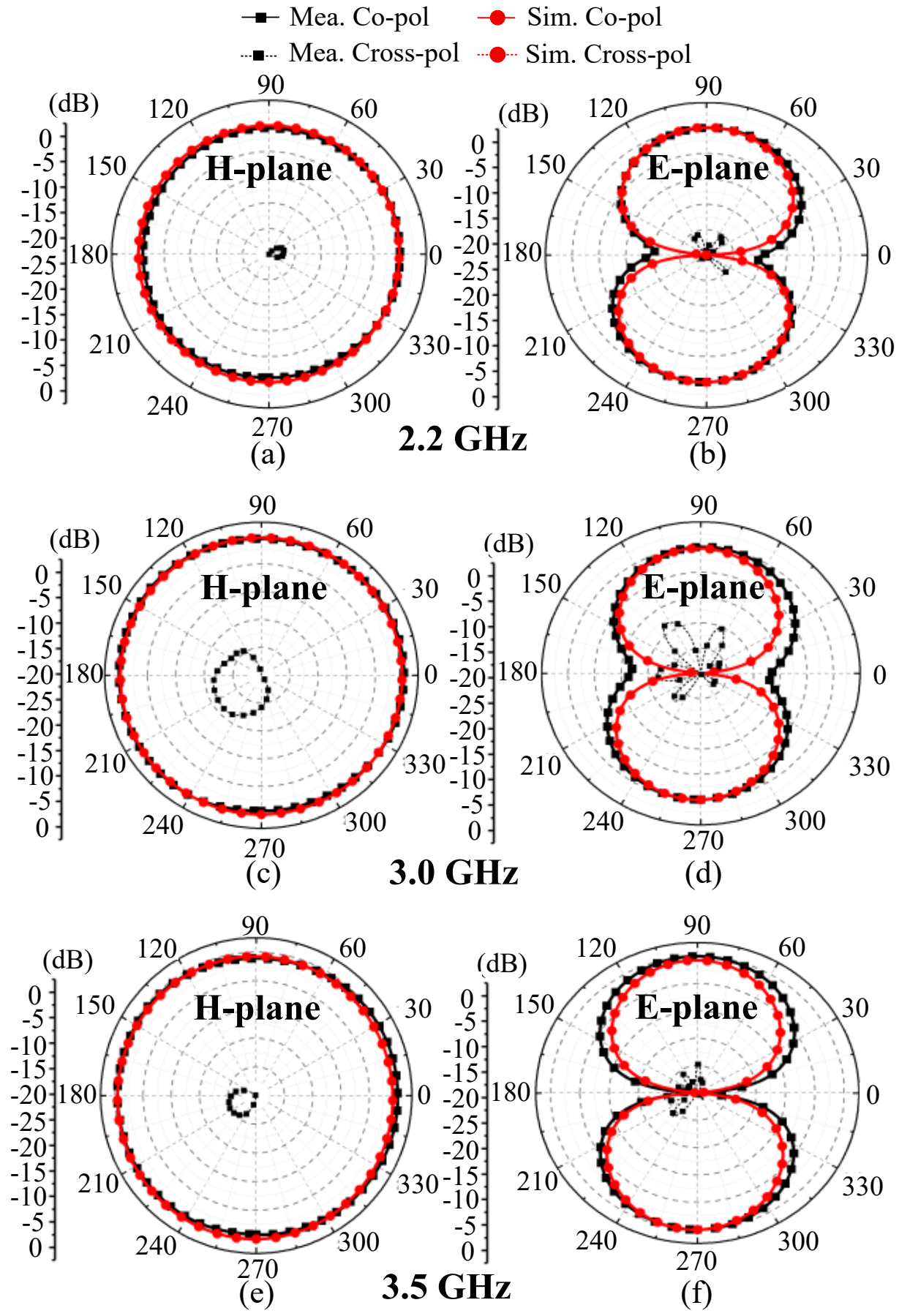

Fig. 29. Measured and simulated radiation patterns of the bent antenna with bending radius ρ = 30 mm.

### B. Comparison and Discussion

In Table I, a comprehensive comparison of the proposed antenna with related published works is summarized. Compared to existing omnidirectional filtering antennas [24]-[31], where the H-plane gain variation is typically less than 2 dB or 3 dB, the proposed design achieves high radiation roundness with H-plane gain variations of less than 0.8 dB in the unbent state and less than 1 dB after bending. Moreover, the bandwidth of the proposed design is further enhanced compared to existing wideband omnidirectional filtering antennas [26]-[28]. Besides, none of the previously reported omnidirectional filtering antennas has flexibility [24]-[31]. The proposed flexible omnidirectional filtering design maintains stable performance even after bending. Furthermore, the proposed design features smaller size, wider bandwidth, and a larger bending curvature tolerance compared with existing flexible filtering antennas [36]-[37]. From the above comparison, the proposed wearable antenna design offers distinct advantages: high omnidirectionality across a wide bandwidth, flexible conformal capability, compact size, and larger working bending curvature.

## IV. Conclusion

In this article, a flexible wideband filtering monopole antenna with symmetric structure for high omnidirectionality is presented. The design is based on a monopole antenna, which is printed on a single-layer flexible substrate. By symmetrically placing a pair of parasitic strips on both sides of the monopole antenna, a radiation null is obtained in the higher band. Furthermore, we introduce a pair of folded parasitic strips on both sides of the feed line along with a slotted metal ground, resulting in a radiation null in the lower band. Besides, the monopole is slotted symmetrically for wideband operation. The proposed antenna underwent fabrication and experimental validation. Measured results reveal that the proposed antenna exhibits high omnidirectional omnidirectionality in the wide operating band due to the implementation of a symmetric filtering structure. Besides, the antenna can maintain stable bandwidth, high omnidirectionality, and effective frequency selectivity under different bending radii. The design has potential applications on various wireless communication systems.